\documentclass[11pt,a4paper,english,twoside]{article}

\usepackage{a4wide}
\usepackage{amssymb, amsmath, amsthm}
\usepackage{graphicx}
\usepackage{subcaption}
\usepackage[all]{xy}
\usepackage{enumerate}
\usepackage[pdftex,hyperref,svgnames]{xcolor}
\usepackage[pdftex,colorlinks=true,
pdfstartview=FitV,
pdfnewwindow=true,
linktoc = page,
linkcolor= Red,
citecolor= blue,
urlcolor= blue,
hyperindex=true,
hyperfigures=false]{hyperref}
\hypersetup{linktocpage}
\usepackage{dsfont}
\usepackage{empheq}
\usepackage{cite}
\usepackage{float}
\usepackage{cancel}
\usepackage{relsize}
\usepackage{soul}
\usepackage{enumitem}
\usepackage{hhline}

\newcommand{\beq}{\begin{equation}}
\newcommand{\eeq}{\end{equation}}
\def\bea#1\eea{\begin{align}#1\end{align}}
\def\beal#1\eeal{\begin{subequations}\begin{align}#1\end{align}\end{subequations}}
\newcommand{\nn}{\nonumber}
\newcommand{\w}{\wedge}
\newcommand{\R}{\mathcal{R}}

\newcommand{\f}[2]{f^{#1}{}_{#2}}

\newcommand{\Rc}{\mathcal{R}}

\newcommand{\ap}{a_{||}}

\newcommand{\api}{a_{||_I}}
\newcommand{\abi}{a_{\bot_I}}
\newcommand{\bp}{b_{||}}

\newcommand{\cp}{c_{||}}

\newcommand{\sun}{\sigma_1}
\newcommand{\sde}{\sigma_2}

\def\del {\partial}
\def\d {{\rm d}}
\def\mmm {\mathcal{M}}

\newcommand{\pb}{\bar{p}}

\def\code#1{\texttt{#1}}

\newtheorem*{lemma}{Lemma}

\begin{document}
\numberwithin{equation}{section}

\begin{titlepage}

\begin{center}

\phantom{DRAFT}

\vspace{2.4cm}

{\LARGE \bf{New de Sitter solutions of 10d type IIB \vspace{0.3cm}\\ supergravity}}\\

\vspace{2.2 cm} {\Large David Andriot$^{1}$, Paul Marconnet$^{2}$, Timm Wrase$^{1,3}$}\\
\vspace{0.9 cm} {\small\slshape $^1$ Institute for Theoretical Physics, TU Wien\\
Wiedner Hauptstrasse 8-10/136, A-1040 Vienna, Austria}\\
\vspace{0.2 cm} {\small\slshape $^2$ Ecole Normale Sup{\'e}rieure de Lyon\\
15 parvis Ren{\'e} Descartes, 69342 Lyon, France}\\
\vspace{0.2 cm} {\small\slshape $^3$ Department of Physics, Lehigh University,\\
16 Memorial Drive East, Bethlehem, PA, 18018, USA}\\
\vspace{0.5cm} {\upshape\ttfamily david.andriot@tuwien.ac.at; paul.marconnet@ens-lyon.fr;\\
timm.wrase@lehigh.edu}\\

\vspace{2.8cm}

{\bf Abstract}
\vspace{0.1cm}
\end{center}

\begin{quotation}
\noindent We find and study 17 new de Sitter solutions of ten-dimensional (10d) type IIB supergravity with intersecting $D_5$-branes and orientifold $O_5$-planes, as well as a new Minkowski one. These solutions are obtained numerically on 6d group manifolds, the compactness of which is established for 4 of them. We show that all our de Sitter solutions are perturbatively unstable, using a restricted 4d effective theory of four scalar fields. We finally analyse whether our solutions can be promoted to classical string backgrounds. Several of them appear as good candidates, as they satisfy all requirements imposed so far.
\end{quotation}

\end{titlepage}

\newpage

\tableofcontents

\section{Introduction}

Cosmological models describing our universe in its present and future state, as well as in its very early stages, exhibit solutions which are close to a pure de Sitter space-time. Observations are nowadays bringing new and tight constraints, that narrow deviations from these models. It is then an important and timely question to ask whether string theory, as a candidate for a fundamental theory of nature, is able to generate a four-dimensional (4d) de Sitter space-time, or slight deviations thereof. As for now, it appears difficult to get from string theory such a de Sitter solution, in a setting where regimes and approximations are well-controlled \cite{Danielsson:2018ztv}. This situation has even led to conjectures \cite{Obied:2018sgi, Andriot:2018wzk, Garg:2018reu, Ooguri:2018wrx, Garg:2018zdg, Andriot:2018mav, Rudelius:2019cfh, Bedroya:2019snp, Andriot:2020lea} in the context of the swampland program, which to various extents, prevent quantum gravity from having (quasi) de Sitter solutions. In string theory, the simplest framework, and thus one having good chances to be well-controlled, is that of classical de Sitter solutions (see e.g.~\cite{Andriot:2019wrs} for a review). These are 10d solutions with a 4d de Sitter space-time times a 6d compact manifold, in a classical regime of string theory, i.e.~specific solutions of a 10d supergravity theory. The simplicity of the setup, allowing only few ingredients, however comes at a price: this option is plagued by many no-go theorems, in agreement with the swampland conjectures, which forbid any de Sitter solution in a large part of parameter space. The remaining part is the one of interest in this paper: we will find there new de Sitter solutions of 10d type IIB supergravity, and we will discuss to what extent they correspond to classical string backgrounds.

Obtaining classical de Sitter solutions in heterotic string has been excluded in \cite{Green:2011cn, Gautason:2012tb, Kutasov:2015eba, Quigley:2015jia}. In the literature, the main focus has thus been on type IIA/B 10d supergravities, with $D_p$-branes and orientifold $O_p$-planes \cite{Maldacena:2000mw, Hertzberg:2007wc, Silverstein:2007ac, Covi:2008ea, Haque:2008jz, Caviezel:2008tf, Flauger:2008ad, Danielsson:2009ff, deCarlos:2009fq, Caviezel:2009tu, Dibitetto:2010rg, Wrase:2010ew, Danielsson:2010bc, Andriot:2010ju, Blaback:2010sj, Danielsson:2011au, Shiu:2011zt, Burgess:2011rv, VanRiet:2011yc, Danielsson:2012by, Danielsson:2012et, Gautason:2013zw, Kallosh:2014oja, Junghans:2016uvg, Andriot:2016xvq, Junghans:2016abx, Andriot:2017jhf, Roupec:2018mbn, Andriot:2018ept, Junghans:2018gdb, Banlaki:2018ayh, Cordova:2018dbb, Cribiori:2019clo, Andriot:2019wrs, Das:2019vnx, Grimm:2019ixq, Cordova:2019cvf, Andriot:2020lea, Kim:2020ysx}. While this framework is the one of interest here, the question of stringy de Sitter solutions has also been tackled recently in various interesting alternatives, including \cite{Terrisse:2019usq, Hardy:2019apu, Dasgupta:2019gcd, deAlwis:2019dkc, Cribiori:2019bfx, Antoniadis:2019rkh, Shukla:2019akv, Cribiori:2019drf, Cribiori:2019hrb, Dibitetto:2019odu, Tsimpis:2020ysl, Sheikhahmadi:2020prj, Bernardo:2020nol, Bento:2020fxj}. Most of the works on classical de Sitter solutions in type II supergravities consider a certain ansatz and setup: the 6d internal space is a group manifold $\mmm$, the fluxes are constant and the $O_p/D_p$ sources are ``smeared'' (see section \ref{sec:ccl} on this last point). In this framework, the no-go theorems on the existence of solutions leave very little possibilities: with $O_p/D_p$ sources of single size $p$, having de Sitter solutions requires $p=4,5$ or $6$, as well as a non-zero $F_{6-p}$ Ramond-Ramond flux. Further constraints of this kind were obtained, such as the need of $\mmm$ to be negatively curved (see \cite{Andriot:2019wrs} for more). The only known 10d supergravity de Sitter solutions obeying this ansatz certainly fall in this small part of parameter space: they were found in type IIA in \cite{Caviezel:2008tf, Flauger:2008ad, Danielsson:2010bc, Danielsson:2011au, Roupec:2018mbn}, with intersecting $O_6/D_6$ and $F_0 \neq 0$. Another, seemingly T-dual de Sitter solution, was found in type IIB with $O_5/O_7$ sources \cite{Caviezel:2009tu}.

This small part of parameter space where such de Sitter solutions are still allowed was explored in \cite{Andriot:2017jhf}, and a strong similarity was noticed between the cases of intersecting $O_6/D_6$ and $O_5/D_5$. In particular, a simplification in the equations would occur if sources share $N_o= p-5$ common internal directions, for $p\geq 5$. This number is also the one allowing the source configuration to preserve some supersymmetry. While this is verified for the known solutions of \cite{Danielsson:2011au}, with $O_6/D_6$ and $N_o=1$, this observation motivates a search for de Sitter solutions in type IIB with intersecting $O_5/D_5$ sources that do not overlap on the 6d manifold, i.e.~$N_o=0$. This is the starting point of this paper, and we display our source configuration in Table \ref{tab:sources}. In addition, as explained in section \ref{sec:Op}, we cannot have $O_5$ along all 6d directions, together with constant fluxes: the orientifold projection would then set $F_1=0$, preventing us from finding de Sitter solutions.
\begin{table}[h]
  \begin{center}
    \begin{tabular}{|c||c|c|c||c|c|c|c|c|c|}
    \hline
Space dimensions & 1 & 2 & 3 & 4 & 5 & 6 & 7 & 8 & 9 \\
    \hhline{==========}
 $O_5, D_5$ & $\otimes$ & $\otimes$ & $\otimes$ & $\otimes$ & $\otimes$ & & & & \\
     \hhline{-||---||------}
 $(O_5), D_5$ & $\otimes$ & $\otimes$ & $\otimes$ & & & $\otimes$ & $\otimes$ & & \\
     \hhline{-||---||------}
 $(D_5)$ & $\otimes$ & $\otimes$ & $\otimes$ & & & & & $\otimes$ & $\otimes$ \\
    \hline
    \end{tabular}
     \caption{$O_5/D_5$ source configuration considered in this paper: they are along the three extended space dimensions, and some of the 6d ones. There is no overlap along the internal dimensions. Also, we do not allow for $O_5$ along all directions. Parentheses indicate that the presence of the source is optional.}\label{tab:sources}
  \end{center}
\end{table}

We detail in section \ref{sec:framework} the ansatz of our solutions and the set of equations and constraints to solve. The numerical procedure used to find solutions is then presented in section \ref{sec:dSsol} and appendix \ref{ap:num}. It allows us to find 17 new de Sitter solutions of type IIB supergravity, with intersecting $O_5/D_5$ on group manifolds, listed explicitly in appendix \ref{ap:sol}, as well as a new Minkowski solution given in section \ref{sec:Mink}. Our method allows for a maximal freedom in the structure constants encoding the group manifold. This has the drawback of making the underlying 6d geometry a little obscure, and in particular, it does not guarantee a priori the compactness of $\mmm$. We discuss this issue in section \ref{sec:compactness} and appendix \ref{ap:alg}, while establishing the compactness in 4 solutions.

All known de Sitter solutions of 10d supergravities with intersecting $O_p/D_p$ were found to be perturbatively unstable. Many stability studies were performed in the literature, either formally or based on concrete examples \cite{Covi:2008ea, Danielsson:2011au, Shiu:2011zt, Danielsson:2012by, Danielsson:2012et, Kallosh:2014oja, Junghans:2016uvg, Junghans:2016abx, Andriot:2018ept, Roupec:2018mbn, Andriot:2019wrs}. In section \ref{sec:pot} and appendix \ref{ap:4d}, we introduce the tools to study the stability of our solutions: a 4d effective theory capturing some scalar fluctuations around our 10d solutions. Building on previous works \cite{Danielsson:2012et, Junghans:2016uvg, Andriot:2018ept, Andriot:2019wrs, Andriot:2020lea}, we consider a 4-field scalar potential $V(\rho,\tau,\sun,\sde)$ and compute the scalar field kinetic terms. This material is sufficient to show in section \ref{sec:stabana} that all our 17 de Sitter solutions are unstable, and we compute in Table \ref{tab:eta} the corresponding $\eta_V$ parameters. More comments and a useful lemma on the mass matrix are given in section \ref{sec:lemma}.

Despite their perturbative instability, it remains crucial to determine whether our de Sitter supergravity solutions correspond as well to classical string backgrounds. Indeed, it is for now unclear that any of the known de Sitter solutions of 10d supergravity achieves this. This question has been recently investigated in various settings \cite{Roupec:2018mbn, Junghans:2018gdb, Banlaki:2018ayh, Andriot:2019wrs, Grimm:2019ixq}, in relation to some swampland conjectures \cite{Ooguri:2018wrx, Bedroya:2019snp} that forbid this possibility in asymptotic limits in field space. We introduce in section \ref{sec:settingquantiz} the requirements to be met by our solutions, as well as 10d tools to test them in this regard. Part of these requirements are then successfully verified by some of our solutions in section \ref{sec:firstcheckquantiz}, where we highlight differences with previous treatments of this matter in the literature. We also indicate limitations in our procedure, related to the absence of a detailed knowledge of the 6d geometry, as mentioned previously. The group manifolds of our solution 14 and 15 are however well identified and understood, so a complete analysis for those will be provided in a companion paper \cite{Andriot:2020vlg}.

More context and references for each of the above topics are provided at the beginning of sections \ref{sec:exist}, \ref{sec:stab} and \ref{sec:regime}, and a summary of our results is given in section \ref{sec:ccl}. Few open questions and future directions are also discussed there.

\section{De Sitter solutions: existence}\label{sec:exist}

In this section, we report on the existence of new de Sitter solutions of 10d type IIB supergravity with intersecting $D_5$ or $O_5$ sources. We first present the mathematical problem to solve in section \ref{sec:framework}, namely the equations and constraints as well as our ansatz for a solution. We then present in section \ref{sec:dSsol} and appendix \ref{ap:num} the procedure used to find such solutions numerically, together with an example of solution found, and further characteristics; the full set of 17 solutions found is given in appendix \ref{ap:sol}. We further discuss the issue of compactness of the 6d group manifold, and we prove the compactness for 4 of our solutions, in section \ref{sec:compactness} and appendix \ref{ap:alg}. We finally present a new Minkowski solution and make further comments in section \ref{sec:Mink}.

\subsection{Setting the stage}\label{sec:framework}

\subsubsection{Solution ansatz}\label{sec:ansatz}

In this paper, we are interested in solutions of 10d type IIB supergravity with $D_5$-branes and orientifold $O_5$-planes as sources. We follow supergravity conventions of \cite{Andriot:2016xvq}, and those of \cite{Andriot:2017jhf} regarding intersecting sources. We consider here a standard solution ansatz presented in \cite{Andriot:2019wrs}, to which we refer for more detail. The 10d space-time is split as a product of a 4d de Sitter space-time, of metric $g_{\mu\nu}$, and a 6d compact group manifold $\mmm$, of metric $g_{mn}$. The 10d metric reads
\beq
\d s_{10}^2 = g_{\mu\nu} \d x^{\mu} \d x^{\nu} + g_{mn} \d y^{m} \d y^{n} \ .
\eeq
We do not include a warp factor, so the sources can be viewed as ``smeared'', or rather, some equations can be considered integrated. We come back in section \ref{sec:ccl} to the question of a localized version of our solutions. The reason for our ansatz is that we will consider intersecting sources, for which a localized description is notoriously difficult to obtain. For the same reason, we take a constant dilaton $e^{\phi}=g_s$. The 6d metric is expressed in a flat basis in terms of 1-forms $e^a$ as follows
\beq
\d s^2_{6}=  g_{mn} \d y^{m} \d y^{n} =  \delta_{ab} e^{a} e^{b} \ ,\quad e^a =e^a{}_m \d y^m\ , \quad \d e^a= -\tfrac{1}{2} f^{a}{}_{bc} e^b\w e^c \ ,
\eeq
where the last equation is the Maurer-Cartan equation. It defines $f^a{}_{bc}$ which will here be taken constant, and thus correspond to structure constants of a Lie algebra. This algebra underlies the group manifold $\mmm$. Compactness of the latter requires $f^a{}_{ac}=0$ (with sum), a condition to be used from now on. The $f^a{}_{bc}$ can be related in full generality to spin connection coefficients (see e.g.~appendix A of \cite{Andriot:2013xca}), so the 6d Ricci tensor in the flat basis can be expressed as
\beq
2\ {\cal R}_{cd} = - f^b{}_{ac} f^a{}_{bd} - \delta^{bg} \delta_{ah} f^h{}_{gc} f^a{}_{bd} + \frac{1}{2} \delta^{ah}\delta^{bj}\delta_{ci}\delta_{dg} f^i{}_{aj} f^g{}_{hb} \ , \label{Ricci}
\eeq
where we specified to a compact group manifold. In the following, we will additionally restrict ourselves to work in a basis of $\{e^a\}$ such that $\f{a}{ac} = 0$ \emph{without sum}, for convenience. This is a priori a further restriction on the ansatz, even though many group manifolds admit such a basis. Finally, in our ansatz, the fluxes are captured by the purely internal 3-form $H$ and 1-, 3-, 5-forms $F_{q=1,3,5}$. We further restrict to constant fluxes, meaning that the flux components in the flat basis are taken constant. With this ansatz, we will see that all entries in the equations to be solved are constant.

In our ansatz, each source $O_5$ or $D_5$ is along the three extended space dimensions, and is wrapping two internal flat directions. For each source, we then split the 1-forms into the two sets $\{ e^{a_{||}} \}$ and $\{ e^{a_{\bot}} \}$, taken globally distinct. Every flat index can then be specified as being parallel or transverse to a given source. For instance, for any internal $q$-form $F_q$, we denote by a label ${}^{(n)}$ its number of legs along a source, with $0 \leq n \leq 2$, meaning
\beq
F_q = \frac{1}{q!} F^{(0)}_{q\, a_{1\bot} \dots a_{q\bot}} e^{a_{1\bot}} \w \dots \w e^{a_{q\bot}} + \frac{1}{(q-1)!} F^{(1)}_{q\, a_{1||} a_{2\bot} \dots a_{q\bot}} e^{a_{1||}} \w e^{a_{2\bot}} \w \dots \w e^{a_{q\bot}} + \dots\ , \label{fluxes}
\eeq
and each $F^{(n)}_{q\, a_1 \dots a_q}$ is here constant. Each source defines naturally parallel and transverse volume forms, ${\rm vol}_{||}$ and ${\rm vol}_{\bot}$, in terms of the $\{ e^{a_{||}} \}$ and $\{ e^{a_{\bot}} \}$. Few more useful conventions on our forms include
\bea
& \epsilon_{1\dots 6}=1 \ ,\ \ {\rm vol}_{||} \w {\rm vol}_{\bot} = {\rm vol}_{6} = \d^{6} y \sqrt{|g_{6}|} = e^1 \w \dots \w e^6 \ ,\\
& *_{6} ( e^{a_1} \w \dots \w e^{a_q}) = \frac{1}{(6-q)!}\, \delta^{a_1b_1} \dots \delta^{a_q b_q} \epsilon_{b_1 \dots b_q c_{q+1} \dots c_6} e^{c_{q+1}} \w \dots \w e^{c_6} \ , \ *_{6}^2 A_q = (-1)^{q} A_q \ , \nn\\
& A_q\w *_{6} A_q = {\rm vol}_{6}\ |A_q|^2 \ , \ |A_q|^2 = A_{q\, a_1\dots a_q}A_{q\, b_1\dots b_q} \delta^{a_1 b_1} \dots \delta^{a_q b_q} /q! \ ,\nn
\eea
and for $p=5$ $O_p/D_p$ sources, one has $*_6 {\rm vol}_{\bot} = {\rm vol}_{||} \ , \ *_6 {\rm vol}_{||} = {\rm vol}_{\bot}$.

In the following, we will consider intersecting sources, and follow notations of \cite{Andriot:2017jhf}. We will have several sets $I=1, \dots, N$ of parallel $O_5/D_5$ that intersect each other. This means that each set wraps a specific pair of internal dimensions. In other words, $\{\{ e^{a_{||_I}} \}, \{ e^{a_{\bot_I}} \}\}$ and $\{\{ e^{a_{||_J}} \}, \{ e^{a_{\bot_J}} \}\}$ are different for $I\neq J$. The above indices $||$ and $\bot$ then get a further label $I$, to specify the set they refer to and corresponding directions. The trace $T_{10}$ of the source energy momentum tensor $T_{MN}$ then gets decomposed into the contributions of each set $I$: $T_{10} = \sum_{I} T_{10}^I$. Each of the $T_{10}^I$ is proportional to $N_s^I=N_{O_5}^I - N_{D_5}^I$, the number of sources in the set $I$, given by the difference of the number of $O_5$ and $D_5$; see \eqref{T10I}. We further restrict ourselves to the case where the sets do not overlap each other, and are orthogonal in the flat basis.\footnote{In general, one would introduce overlap numbers $\delta_{a_{||_I}}^{a_{||_J}}$, indicating the number of common directions between the two sets $I,J$ \cite{Andriot:2017jhf}. We restrict to the case of homogeneous overlap where there is only one number $\ \forall I,\, J\neq I,\ \delta_{a_{||_I}}^{a_{||_J}} = N_o$, and take $N_o=0$. This matches the natural number $N_o= p-5$ for $O_p/D_p$ \cite{Andriot:2017jhf} as mentioned in the Introduction.} This choice leaves two possibilities: $N=2$ or $N=3$. The former can be studied through the latter by setting $T_{10}^{I=3}=0$, and we will do so. However, whether $O_5$ are present in each set or not makes a difference, and we will come back to this point. Without loss of generality, we then place the $N=3$ sets along internal flat directions (12), (34), (56), i.e.~defining the following volume forms
\beq
\begin{aligned}
	\label{sets}
& I=1:\quad {\rm vol}_{||_1} = e^1 \w e^2 \ ,\ {\rm vol}_{\bot_1} = e^3 \w e^4 \w e^5 \w e^6 \ ,\\
& I=2:\quad {\rm vol}_{||_2} = e^3 \w e^4 \ ,\ {\rm vol}_{\bot_2} = e^1 \w e^2 \w e^5 \w e^6 \ ,\\
& I=3:\quad {\rm vol}_{||_3} = e^5 \w e^6 \ ,\ {\rm vol}_{\bot_3} = e^1 \w e^2 \w e^3 \w e^4 \ .
\end{aligned}
\eeq
As explained in the following, the set $I=1$ will contain $O_5$, the set $I=2$ may contain some as well, while the set $I=3$ will not.

\subsubsection{Orientifold projection}\label{sec:Op}

Having $O_5$ requires to impose the orientifold projection, leading to important restrictions. For each $O_5$, the only possible non-zero structure constants are the following
\beq
f^{a_{||}}{}_{b_{\bot}c_{\bot}},\ f^{a_{\bot}}{}_{b_{\bot}c_{||}},\ f^{a_{||}}{}_{b_{||}c_{||}} \ .\label{fabcOp}
\eeq
The choice of working in a basis where $\f{a}{ac} = 0$ without sum implies for an $O_5$ that $\f{\ap}{\bp \cp} = 0$, leaving us with only two types of structure constants. The components \eqref{fluxes} of the fluxes are also limited by the projection to the following
\beq
O_5: \quad F_1^{(0)}, F_3^{(1)}, F_5^{(2)}, \quad H^{(0)}, H^{(2)} \label{fluxcompo} \ .
\eeq
In addition, $F_5=F_5^{(2)}$ implies $*_6 F_5 = (*_6 F_5)^{(0)}$, entering the equations. Finally, one has to impose these restrictions for the $O_5$ present in each source set $I$. This leads to an important observation: if $O_5$ are present in each of the $N=3$ sets, then any internal direction is parallel to one $O_5$. This implies that $F_1=F_1^{(0)}$, by definition a purely transverse form with constant component, has to vanish. This flux is however mandatory to get de Sitter solutions with intersecting $p=5$ sources (see e.g.~\cite{Andriot:2019wrs}). As mentioned in the Introduction, we conclude that one cannot have $N=3$ with $O_5$ along each set. Rather, we will have one set with only $D_5$. We still need to have $O_5$ \cite{Maldacena:2000mw} (in the case of intersecting sources, this is reflected in $T_{10}>0$, while the $T_{10}^{I}$ can be of different signs \cite{Andriot:2017jhf}). In short, our set $I=1$ will always contain $O_5$, the set $I=2$ may contain some, and the set $I=3$ does not. This is implemented in the constraint $T_{10}^3 \leq 0$.\footnote{Requiring a de Sitter solution through $\Rc_4 >0$ implies that $T_{10}>0$, as can be seen e.g.~in \eqref{R4T_{10}F}, giving $T_{10}^{1}>0$  or $T_{10}^{2}>0$. If $T_{10}^{1}<0$, one can still have $O_5$ in the set $I=1$, their contribution is simply dominated by that of $D_5$. We then do not need to impose more constraint. In practice, all our solutions will have $T_{10}^{1}>0$.}

Even though we may only have $O_5$ in one set along (12), we now impose for simplicity the projection for possible $O_5$ in sets $I=1$ and $I=2$. A first projection along (12) of one $O_5$ keeps 28 flux components and 36 structure constants. Out of those, the second one along directions (34) only leaves the following variables
\bea
F_1:&\quad F_{1 \ 5} \,, \quad F_{1 \ 6} \,,\nn\\
F_3:&\quad F_{3\ 315} \,, \quad F_{3\ 316} \,, \quad F_{3\ 325} \,, \quad F_{3\ 326} \,, \quad F_{3\ 415} \,, \quad F_{3\ 416} \,, \quad F_{3\ 425} \,, \quad F_{3\ 426} \,,\nn\\
F_5:&\quad F_{5 \ 34125} \,, \quad F_{5 \ 34126} \,, \nn\\
H:&\quad H_{125} \,, \quad H_{126} \,, \quad H_{345} \,, \quad H_{346} \,,\label{variables}\\
\f{a_{||_2}}{b_{\bot_2} c_{\bot_2}}:&\quad \f{3}{15} \,, \quad \f{3}{16} \,, \quad \f{3}{25} \,, \quad \f{3}{26} \,, \quad \f{4}{15} \,, \quad \f{4}{16} \,, \quad \f{4}{25} \,, \quad \f{4}{26} \,,\nn\\
\f{a_{\bot_2}}{b_{\bot_2}c_{||_2}}:&\quad \f{1}{53} \,, \quad \f{1}{63} \,, \quad \f{1}{54} \,, \quad \f{1}{64} \,, \quad\f{2}{53} \,, \quad \f{2}{63} \,, \quad \f{2}{54} \,, \quad \f{2}{64} \,, \nn\\
&\quad\f{5}{13} \,, \quad \f{5}{23} \,, \quad \f{5}{14} \,, \quad \f{5}{24} \,, \quad\f{6}{13} \,, \quad \f{6}{23} \,, \quad \f{6}{14} \,, \quad \f{6}{24} \,,\nn
\eea
where the structure constants could equivalently be classified according to the set $I=1$. The second projection reduces the number of independent variables to 16 fluxes and 24 structure constants. With the 3 source contributions $T_{10}^I$, this adds up to 43 variables. Those will enter the equations to be solved, that we now detail.

\subsubsection{Equations}\label{sec:eq}

Given the ansatz for de Sitter solutions with $O_5/D_5$ ($p=5$ in the following) presented in section \ref{sec:ansatz}, the type IIB supergravity equations to solve, in 10d string frame, are the following equations of motion (e.o.m.) and Bianchi identities (BI)
\begin{itemize}
\item the fluxes e.o.m.
\bea
& \d ( *_6 H) - g_s^2 ( F_{1} \w *_6 F_{3} + F_{3} \w *_6 F_{5} )  = 0 \ ,\\
& \d( *_6 F_1 ) + H \w *_6 F_{3} = 0\ , \label{F_1eom} \\
& \d( *_6 F_3 ) + H \w *_6 F_{5} = 0\ ,\\
&\d ( *_6 F_5 ) = 0 \ ,
\eea
\item the fluxes BI
\bea
& \d H =0 \ ,\\
& \d F_1=0 \ ,\\
& \d F_3 - H \w F_1 = - \sum_{I} \frac{T_{10}^I}{p+1} \, {\rm vol}_{\bot_I}  \ , \label{BI2}\\
& \d F_5 - H \w F_3 = 0 \ , \label{F5BI}
\eea
\item the dilaton e.o.m.
\beq
2 {\cal R}_{4}+ 2{\cal R}_6 + g_s \frac{T_{10}}{p+1} -|H|^2 = 0 \ , \label{dileom}
\eeq
\item the 4d Einstein equation (equivalent to its trace)
\beq
4 {\cal R}_4 = 2 g_s \frac{T_{10}}{p+1} - 2|H|^2 - g_s^2 ( 2 |F_3|^2 + 4 |F_5|^2 ) \ ,
\eeq
\item the 6d (trace-reversed) Einstein equation
\bea
{\cal R}_{ab} & = \frac{g_s^2}{2}\left(F_{1\ a}F_{1\ b} +\frac{1}{2!} F_{3\ acd}F_{3\ b}^{\ \ \ cd} + \frac{1}{2 \cdot 4!} F_{5\ acdef}F_{5\ b}^{\ \ \ cdef} - \frac{1}{2} *_6 F_{5\ a} *_6 F_{5\ b} \right) \nn\\
& \ + \frac{1}{4} H_{acd}H_b^{\ \ cd} + \frac{g_s}{2}T_{ab} + \frac{\delta_{ab}}{16} \left( - g_s T_{10} - 2|H|^2 - 2 g_s^2 |F_3|^2 \right)  \ ,\label{6dEinstein}\\
{\rm with} &\ \ T_{ab} = \sum_I \delta^{a_{||_I}}_{a} \delta^{b_{||_I}}_{b} \delta_{a_{||_I}b_{||_I}} \frac{T_{10}^I}{p+1} \ ,\label{Tab}
\eea
\item the Riemann BI or Jacobi identity
\beq
f^a{}_{e[b} f^e{}_{cd]}=0 \ .
\eeq
\end{itemize}
On group manifolds, the Riemann BI is indeed equivalent to the Jacobi identity of the algebra, see e.g.~(3.5) of \cite{Andriot:2014uda}. Finally, to guarantee the validity of the solution, one should also check
\begin{itemize}
  \item Additional requirements:
  \beq
  \text{de Sitter:}\ {\cal R}_4 > 0 \ ,\quad \text{orientifold projection(s)} \ ,\quad \text{compactness of}\ \mmm \ .
  \eeq
\end{itemize}
The orientifold projection has been imposed in section \ref{sec:Op} by selecting the non-zero flux components and structure constants, up to the requirement $T_{10}^3 \leq 0$, consistent with the placement of the sources. In other words, using the variables \eqref{variables} makes these projections satisfied, and we will do so when looking for solutions. Ensuring the compactness of $\mmm$ amounts to identify the underlying algebra and manifold, and verify the existence of a lattice. This is a non-trivial task that we will discuss in detail in section \ref{sec:compactness}. Once all these equations and constraints are solved, one may discuss the consistency of such a supergravity solution as a classical string background: we turn to this question in section \ref{sec:regime}.

The trace of the 6d Einstein equation combined with above equations leads to the following useful expression \cite{Andriot:2017jhf}
\beq
{\cal R}_4= g_s \frac{T_{10}}{p+1} - g_s^2 \sum_{q=1,3,5} |F_q|^2 \label{R4T_{10}F} \ .
\eeq
It can be traded for one of the ${\cal R}_4$ expressions above. This equation provides the requirement of having $T_{10}>0$ \cite{Maldacena:2000mw} also for intersecting sources, i.e.~the need here of some orientifold.\\

With the ansatz of section \ref{sec:ansatz}, several simplifications occur in the above equations. To start with, the e.o.m. for $F_1$ \eqref{F_1eom} and the BI for $F_5$ \eqref{F5BI} are trivially satisfied: the left-hand side are both 6-forms which on the one hand are odd under $O_5$ projections, while on the other hand, they are proportional to the form $\text{vol}_6$ which is even, given the fluxes are constants. So these forms vanish identically. We also get simplifications in the 6d Einstein equations. Indeed, for any set $I$, one can decompose the flat indices into the basis $a_{||_I}, a_{\bot_I}$. The internal Einstein equation can be decomposed into parallel components ${}_{a_{||}b_{||}}$, transverse ones ${}_{a_{\bot}b_{\bot}}$, and ``off-diagonal'' ones ${}_{a_{||}b_{\bot}}$, for each set. For a set where there is an $O_5$, the projection imposes important constraints. This reasoning was presented at the beginning of section 3.2 of \cite{Andriot:2019wrs} for parallel sources, and it still holds here with intersecting ones, the key point being that the source term \eqref{Tab} remains here diagonal
\beq
\hspace{-0.15in}\text{For our sets \eqref{sets} of $O_5/D_5$}:\quad T_{ab} = \text{diag}\left(\! \frac{T_{10}^1}{p+1}, \frac{T_{10}^1}{p+1}, \frac{T_{10}^2}{p+1},\frac{T_{10}^2}{p+1},\frac{T_{10}^3}{p+1},\frac{T_{10}^3}{p+1} \!\right) \, ,
\eeq
with $p=5$. The consequence is that the off-diagonal ${}_{a_{||_I}b_{\bot_I}}$ Einstein equations for $I$ with an $O_5$ projection are trivially satisfied. With an $O_5$ along $I=1,2$, or equivalently using the variables \eqref{variables}, we are left with only the diagonal blocks of the Einstein equations, i.e.~9 equations. In the next subsection, we turn to solving this whole set of equations.\\

Before doing so, let us briefly consider the impact on the above equations of including anti-$D_5$-branes ($\bar{D}_5$) to our sets of sources. We denote the contribution of $\bar{D}_5$ to the set $I$ by $\bar{T}_{10}^I$. As a convention, in a case without $O_5$ and where the distribution of $D_5$ is precisely the same as that of $\bar{D}_5$, one has $\bar{T}_{10}^I=T_{10}^I$. This is consistent with the fact that an $\bar{D}_5$ has the same tension as a $D_5$, and the source energy momentum tensor comes from the DBI action which carries the tension. We thus have $\bar{T}_{10}^I \leq 0$ with conventions of \cite{Andriot:2016xvq}. Among the above equations, the dilaton and Einstein equations are then formally unchanged provided
\beq
T_{10}= \sum_I T_{10}^I + \bar{T}_{10}^I \ ,\quad T_{ab} = \sum_I \delta^{a_{||_I}}_{a} \delta^{b_{||_I}}_{b} \delta_{a_{||_I}b_{||_I}} \frac{T_{10}^I+\bar{T}_{10}^I}{p+1} \ .
\eeq
On the contrary, an $\bar{D}_5$ has opposite charge with respect to a $D_5$, which affects the WZ source action, and thus modifies the Bianchi identity \eqref{BI2}. It gets rewritten as
\beq
\d F_3 - H \w F_1 = - \sum_{I} \frac{T_{10}^I - \bar{T}_{10}^I}{p+1} \, {\rm vol}_{\bot_I}  \ . \label{BI2antiD5}
\eeq
Without considering $\bar{D}_5$ any further, we are now going to solve the equations.

\subsection{De Sitter solutions}\label{sec:dSsol}

We are looking for de Sitter solutions of 10d type IIB supergravity with intersecting $O_5/D_5$ sources. We have presented a solution ansatz in section \ref{sec:ansatz} and \ref{sec:Op}. The problem then amounts to solving a large system of equations given in section \ref{sec:eq}, subject to the constraints
\begin{equation}
\label{constraints}
\Rc_4 > 0 \,, \qquad T_{10}^3 \leq 0 \ ,
\end{equation}
where the last condition is related to the placement of $O_5$ and their projection. Also, we treat $\Rc_4$ as a combination of the variables that should have a definite sign. So the equations depend on 43 variables, the flux components and structure constants of \eqref{variables}, allowed by the orientifold projections, and the three source contributions $T_{10}^I$. In components, this reduces to a set of 46 scalar equations. Although they are at most quadratic in the variables, this remains a computationally demanding problem to solve.

To find solutions, we develop a numerical procedure presented in appendix \ref{ap:num}. It allows us to find numerical solutions to a very good accuracy: the equations are typically satisfied up to a typical maximal error $\varepsilon \sim 10^{-15}$. This should be compared to the value of $\Rc_4$ that would always fall into the interval $[10^{-3}, 10^{-1}]$ (making it clear that we are not finding a Minkowski solution), or the value of the variables in the solutions, that are always in the range $[10^{-4}, 10]$. These values are reasonably large compared to the numerical error. The accuracy could also be checked against a case of a no-go theorem, e.g.~$F_1=0$, for which the error could not be made lower than $\varepsilon \sim 10^{-5}$.

When looking for solutions, we usually obtain a non-zero value for most of the variables. A next step is then to look for simpler solutions, where several variables are either vanishing or related to one another. It is especially important to reduce the number of non-zero structure constants, to help identifying the underlying algebra and verify the compactness of the internal space, as we will discuss in section \ref{sec:compactness}. Starting with a general solution, we then incrementally set the variables with smaller value to zero, while checking that one still has a good solution. Further educated guesses allow us to eventually reduce considerably the number of non-zero variables in our solutions.\\

With this procedure, we obtain 17 de Sitter solutions, that we list explicitly in appendix \ref{ap:sol}. Let us give here one example: solution 16. The values of the variables have 16 significant digits, but we round them here for the sake of readability. The non-zero variables take the following values
\bea
\f{2}{35} & = -0.35847, \quad \f{2}{45} = 0.95728, \quad  \f{2}{46} = -0.59118, \quad \f{3}{15} = 0.21904, \f{3}{16} = 0.18899, \nn\\
\f{4}{15} & = 0.11460, \quad \f{6}{14} = -0.045686, \quad \f{3}{25} = -\f{4}{15}, \quad \f{1}{45} = -\f{2}{35}, \quad g_s F_{1\ 5} = -0.38308, \nn\\
g_s F_{3\ 136} &= 0.35228, \quad g_s F_{3\ 235} = 0.50883, \quad g_s F_{3\ 236} = 1.0454, \quad F_{3\ 246} = F_{3\ 136}, \nn \\
H_{125} &= 0.039232, \quad H_{126} = -0.093956, \quad H_{345} = -0.012542, \quad H_{346} = 0.29391,\nn\\
g_s T_{10}^1 & = 10,\quad g_s T_{10}^2 =1.0654,\quad g_s T_{10}^3 = -0.28655. \label{solex}
\eea
For this solution, we have $\Rc_4 = 0.049845$ and $\varepsilon \sim 10^{-16}$.

More generally, all our 17 solutions have a vanishing $F_5$, and a non-zero $T_{10}^3$. In addition, we managed to set to zero some structure constants in 10 solutions. The solution 14 is very special for several reasons, one being that it is the only solution with $T_{10}^2 < 0$. With only $T_{10}^1 > 0$, this solution falls into the small subset described in section 4.4 of \cite{Andriot:2019wrs}, that is very constrained. We verify in particular for this solution 14 the constraint (4.30) of \cite{Andriot:2019wrs} by computing the quantity
\beq
\lambda_1 = -\frac{\delta^{cd}   f^{b_{\bot_1}}{}_{a_{||_1}  c_{\bot_1}} f^{a_{||_1}}{}_{ b_{\bot_1} d_{\bot_1}} }{\tfrac{1}{2} \delta^{ab} \delta^{cd} \delta_{i j}  f^{i_{||_1}}{}_{a_{\bot_1} c_{\bot_1}} f^{j_{||_1}}{}_{b_{\bot_1} d_{\bot_1}}} = 0.0020380 \ ,\label{lambda1}
\eeq
which is indeed between 0 and 1 as required there.

Finally, as detailed in appendix \ref{ap:num}, there seems to be no solution with $T_{10}^2 = 0 = T_{10}^3$, which would correspond to a solution with parallel sources. This is in agreement with conjecture 1 of \cite{Andriot:2019wrs}. Also, there seems to be no solution with only 1 or 2 non-vanishing structure constants. In comparison, the smallest number of non-zero $\f{a}{bc}$ found is 7 (at least in this basis), in solution 15.\\

As mentioned in section \ref{sec:eq}, for a given solution, we are left to check the compactness of the 6d internal manifold $\mmm$, through the existence of a lattice. We turn to this task in the next section.

\subsection{Compactness and basis choice}\label{sec:compactness}

The solutions have been searched on 6d group manifolds, defined by a set of structure constants corresponding to an underlying Lie algebra. To make sense of our solutions in a compactification context, we need to identify each of these group manifolds, and verify that they are compact. To that end, one should check that the group manifold admits a lattice, i.e.~a discrete subgroup that provides discrete identifications of the coordinates allowing to make it compact. For instance, a circle can be viewed as the non-compact group $(\mathbb{R},+)$ divided by the lattice $\mathbb{Z}$, the coordinate identification being then $x \sim x+1$. The existence of a lattice is not always guaranteed: see \cite{Andriot:2010ju, Danielsson:2011au} for reviews.

Identifying the group manifolds and verifying the existence of a lattice first requires to identify the underlying algebra. Lie algebras, through Levi decomposition, split as a semi-direct sum into semi-simple algebras and solvable algebras. The most general algebras can be a mixture of both, see e.g.~examples in \cite{Danielsson:2011au}, and we will restrict here ourselves for simplicity to the solvable ones. For example, nilpotent algebras, or almost-abelian solvable algebras, two subsets of solvable algebras, are known to admit lattices \cite{Andriot:2010ju}. Lattices have been shown to exist or to be excluded for further instances of solvable algebras in e.g.~\cite{Bock}. So we should first identify the underlying algebra.

The difficulty in doing so is that algebras are defined up to isomorphisms, for instance relabelings of directions or other change of basis. In our search for solutions, we have let the structure constants be free variables (subject to the Jacobi identities). The directions were from the start made appropriate to the sources, i.e.~parallel or transverse to them. This is reflected in the fact that our metric was simply $\delta_{ab}$ in that basis. It is however possible to pick another basis, less convenient with respect to sources directions, where the metric is more involved, but where the algebra appears much simpler, in particular exhibiting less structure constants. It is typically in such a basis that the (isomorphism class of) algebra is given in classification tables, as those of \cite{Andriot:2010ju} or \cite{Bock}. In addition, for illustration, there exist 164 indecomposable six-dimensional solvable algebras, including 24 nilpotent ones. It is thus not a simple task to identify our algebras within the classified ones.

Fortunately, some properties are inherent to the algebra, i.e.~basis independent. The first one is whether it is solvable, and in that case, what is its nilradical: see \cite{Andriot:2010ju} for definitions. These are simple properties that can be determined given the set of non-zero structure constants. We thus verify that our solutions 1 to 13 are not on solvable algebras. One reason is certainly the high number of structure constants, which probably hints at a mixture of semi-simple and solvable; we refrain from identifying those. Solutions 14 to 17 are on solvable, non-nilpotent algebras. To find them in the algebra classification of \cite{Bock}, we further identify their nilradical. We give details on these identifications in appendix \ref{ap:alg}, and summarize here our results.\\

Solutions 16 and 17 admit as nilradical a five-dimensional, indecomposable, two-step nilpotent algebra, identified as $\mathfrak{g}_{5.3}$. This allows us to further identify the algebra for these two solutions as being $\mathfrak{g}_{6.76}^{-1}$ in table 27 of \cite{Bock}. Indeed, we determine explicitly an isomorphism for each algebra of these solutions to the algebra $\mathfrak{g}_{6.76}^{-1}$. According to Theorem 8.3.4 of \cite{Bock} and the following remark there, this algebra admits a lattice. We conclude that the group manifold for these two solutions can be made compact. The identified algebra remains complicated, as well as the details of its lattice, so the corresponding geometry of the group manifold is not easy to describe.\footnote{It is in addition not guaranteed that one-forms in that basis are globally defined: see \cite{Andriot:2010ju} for a discussion on this. Having globally defined forms may require a further change of basis, not necessarily simple here.} We will then not focus more on the geometry of $\mmm$ for these two solutions.

We proceed similarly for solutions 14 and 15, which turn out to be much simpler. Their nilradical is the four-dimensional abelian algebra, denoted $\mathfrak{n} = 4 \mathfrak{g}_1$. With changes of basis, we can bring the algebras of both solutions to have only four structure constants. It is then easy to see that both algebras are decomposable, into two three-dimensional solvable algebras, each of nilradical $2\mathfrak{g}_1$. For solution 15, we identify the algebra as being $\mathfrak{g}_{3.4}^{-1} \oplus \mathfrak{g}_{3.4}^{-1}$, and for solution 14, we get $\mathfrak{g}_{3.5}^{0} \oplus \mathfrak{g}_{3.5}^{0}$. All of those admit lattices, so the group manifolds of these two solutions can be made compact. This time, their geometry is simple, and we will come back to them.

To conclude, for 4 out of 17 solutions, lattices could be found so the manifold can be made compact. This ends the validity checks of these de Sitter solutions of type IIB supergravity. For the remaining 13 solutions, we do not know for now. Let us emphasize once more the role of the choice of basis: our choice provided the simple metric $\delta_{ab}$ and directions appropriate to the sources. The change of basis or isomorphisms considered above and in appendix \ref{ap:alg} reduce the number of structure constants, but also act on the metric, generating off-diagonal terms (see \eqref{g-1}). The initial freedom in choosing the structure constants and setting the sources directions eventually corresponds to a freedom in a generic $6\times 6$ metric $g_{ab}$; the simplicity gained in having fewer structure constants is traded for the initial simplicity of the metric $\delta_{ab}$. From this perspective, it is thus unclear whether one basis is simpler when searching for solutions.

\subsection{Minkowski solutions}\label{sec:Mink}

While looking for de Sitter solutions, we encountered accidentally few Minkowski solutions. Most of them had a single set of sources, and fell into the class of \cite{Andriot:2016ufg}, but one Minkowski solution found had two intersecting sets, both containing $O_5$. The list of known Minkowski solutions with intersecting sources on group manifolds is short and given in section 5 of \cite{Andriot:2017jhf}. So the one found here is new, to the best of our knowledge. It is given as follows
\bea
\f{2}{35} &= -0.39104,  \quad \f{4}{16} = 1.3741,\nn\\
g_s F_{1\ 5} &= 1, \quad g_s F_{1\ 6} = -0.39696,\quad g_s F_{3\ 245} = -1.3897,\quad g_s	F_{3\ 246} = -0.33164, \label{Minksol}\\
H_{125} &= 0.23785, \quad H_{126} = -0.84691, \quad H_{345} = -0.24101, \quad H_{346} = -0.57067, \nn\\
g_s T_{10}^1 & = 3.2199,\quad g_s T_{10}^2 = 15.972,\quad g_s T_{10}^3 = 0.\nn
\eea
The 6d manifold is ${\rm Nil}_3 \oplus {\rm Nil}_3$, the direct sum of twice the three-dimensional nilmanifold ${\rm Nil}_3$. One is along directions 235 (with fiber 2) and the other 416 (with fiber 4). The sources wrap directions 12, and 34, thus going across these two subspaces. Given that this solution is not listed in \cite{Andriot:2017jhf} and references therein, it is likely not to be supersymmetric; it would be interesting to verify this point.

Studying possible relations between de Sitter and Minkowski solutions is interesting: indeed, if the latter can be obtained as a limit of the former, this can have implications for stability, as shown e.g.~in theorems like those of \cite{Kallosh:2014oja, Junghans:2016abx}. According to the list of known solutions of section 5 in \cite{Andriot:2017jhf}, no Minkowski solution with intersecting sources has been found on the group manifolds identified in section \ref{sec:compactness} for our de Sitter solutions. There is one possible exception of a fluxless, i.e.~purely geometric solution on $\mathfrak{g}_{3.5}^{0} \oplus \mathfrak{g}_{3.5}^{0}$, since the latter can be made Ricci flat: see section 2.4 of \cite{Andriot:2015sia}. Apart from this, the above de Sitter solutions thus appear so far isolated. However, the new Minkowski solution \eqref{Minksol} could correspond to a limit of the de Sitter solution 14 on $\mathfrak{g}_{3.5}^{0} \oplus \mathfrak{g}_{3.5}^{0}$ and solution 15 on $\mathfrak{g}_{3.4}^{-1} \oplus \mathfrak{g}_{3.4}^{-1}$. Indeed, by setting to 0 two of the four structure constants in those de Sitter solutions, one goes from these solvmanifolds to the nilmanifolds of the Minkowski solution. Taking that limit can either be done by setting directly to 0 the number in the structure constant, or sending to infinity a ratio of radii entering there (see section \ref{sec:regime}). These limits can nevertheless not be viewed strictly speaking as smooth limits, since one eventually changes the manifold topology. To be sure that the Minkowski solution \eqref{Minksol} corresponds to such limits of the de Sitter solutions found, one should further analyse the flux components as well as the sources contributions, and we leave this to future work.

\section{De Sitter solutions: stability}\label{sec:stab}

In this section, we analyse the stability of the de Sitter solutions presented in section \ref{sec:dSsol}. As mentioned in the Introduction, all known de Sitter solutions of type II supergravities with intersecting $O_p/D_p$ sources have been found classically unstable. To show this, one should study fluctuations around a given solution. This is typically done using a 4d effective theory with scalar fields, and studying the scalar potential. The de Sitter solution is then a critical point of this potential, and the instability corresponds to this point being a maximum along one (tachyonic) field direction. The works \cite{Covi:2008ea, Kallosh:2014oja, Junghans:2016uvg, Junghans:2016abx} have provided a better understanding of what appears to be a systematic tachyon in these de Sitter solutions. The focus was on the case where a de Sitter solution is close to a no-scale Minkowski one of 4d ${\cal N} =1$ supergravity. The tachyon would then align with the sgoldstino direction in the Minkowski limit. In spite of these interesting results, and others detailed below, it remains unclear whether a tachyon is indeed present in all possible classical de Sitter solutions (see conjecture 2 of \cite{Andriot:2019wrs}). This motivates us to test the stability of the new de Sitter solutions obtained in this paper.

For previously known de Sitter solutions, a full ${\cal N} =1$ 4d supergravity theory and its scalar potential have been used to analyse the stability. Given the solutions were found on group manifolds with constant fluxes and smeared sources, the 4d gauged supergravity used was most likely a consistent truncation of the 10d theory. Here, we do not have at hand the analogous 4d supergravity that would correspond to our 10d setting with intersecting $O_5/D_5$, even though it may exist in the literature. Instead, we will proceed with a more drastic, though standard, truncation, where we only keep 4 scalar fields and freeze any other. The analysis is simpler, the relation to 10d is straightforward, and if a tachyonic mode is found within these few fields, it is sufficient to prove an instability; we come back to this point in section \ref{sec:lemma}.

The volume $\rho$ and 4d dilaton $\tau$ are well-known 4d scalar fields first introduced in \cite{Hertzberg:2007wc}. In \cite{Danielsson:2012et}, it was proposed to consider a third one, $\sigma$, distinguishing the internal volumes parallel and transverse to the sources. In addition, the tachyon was proposed to lie among these few scalars. This idea was successfully checked on some examples in \cite{Danielsson:2012et, Junghans:2016uvg}. The full scalar potential $V(\rho, \tau, \sigma)$ was worked-out for parallel sources in \cite{Andriot:2018ept} and $V(\rho, \tau, \sigma_I)$ for intersecting sources in \cite{Andriot:2019wrs}; see also \cite{Andriot:2020lea} for an overview and a proper derivation of the $F_5, F_6$ terms which are more subtle. This potential was further used in an attempt to formally prove the presence of a systematic tachyon among these scalar fields \cite{Andriot:2018ept, Andriot:2019wrs}. We now present in section \ref{sec:pot} the 4-field potential $V(\rho, \tau, \sun, \sde)$ to be used, as well as the kinetic terms for these scalars, that are computed in detail in appendix \ref{ap:4d}. We use this material in section \ref{sec:stabana} to show that all our 17 solutions are tachyonic. We compute the corresponding parameter $\eta_V$ and summarize the results in Table \ref{tab:eta}. We finally comment in section \ref{sec:lemma} on the impact on the mass matrix of including more fields in a theory, thanks to a useful lemma.

\subsection{The 4d scalar potential and kinetic terms}\label{sec:pot}

Starting with 10d type IIA/B supergravity action with $O_p/D_p$ sources, one can consider scalar fluctuations around background valued 10d fields, denoted with ${}^0$ when necessary. The background will be for us the above de Sitter solutions, whose ansatz was given in section \ref{sec:ansatz}. Introducing the scalar fluctuations in the 10d action, and integrating over the 6d compact manifold, one obtains a 4d theory for these 4d scalars coupled to gravity. Going to 4d Einstein frame, one eventually obtains the 4d action
\begin{equation}
{\cal S} = \int \d^4 x \sqrt{|g_4|} \left(\frac{M_p^2}{2} \Rc_4 - \frac{1}{2} g_{ij} \del_{\mu}\phi^i \del^{\mu}\phi^j - V \right) \ ,\label{S4dgen}
\end{equation}
with the field space metric $g_{ij}$, a scalar potential $V$ depending on the scalar fields $\phi^i$, and the 4d reduced Planck mass $M_p$ given here by
\beq
M_p^2 = \frac{1}{\kappa_{10}^2} \int \d^6 y \sqrt{|g_6^0|}\ g_s^{-2} \ . \label{Mp}
\eeq
The convention for $M_p$ differs by a factor of $2$ with respect to \cite{Andriot:2018ept}. In the following and in appendix \ref{ap:4d}, we briefly discuss the derivation of the 4d action \eqref{S4dgen}, i.e.~that of the scalar potential $V$ and the kinetic terms. We refer to section 4 of \cite{Andriot:2019wrs} for details, or to \cite{Andriot:2020lea} for an overview, and only focus here on few points specific to our setting.\\

The scalar fields $\phi^i$ are $\rho, \tau, \sigma_{I=1,2,3}$, obtained by fluctuating the 6d metric and the 10d dilaton. By definition, their background value is $\rho = \tau = \sigma_{I} = 1$. Given our sets of sources \eqref{sets}, the metric fluctuations read as follows on the 1-forms
\begin{equation}
e^{1,2}=\sqrt{\rho \sigma_1^A \sigma_2^B \sigma_3^B }\, (e^{1,2})^0 \,, \quad
e^{3,4}=\sqrt{\rho \sigma_1^B \sigma_2^A \sigma_3^B }\, (e^{3,4})^0 \,, \quad
e^{5,6}=\sqrt{\rho \sigma_1^B \sigma_2^B \sigma_3^A }\, (e^{5,6})^0 \ ,
\end{equation}
where $A=p-9, B=p-3$, i.e.~here $p=5$ and $A=-4, B=2$. As observed already in \cite{Danielsson:2012et, Andriot:2019wrs} for $O_6/D_6$, one of the fluctuations is in fact redundant. Indeed, one can set $\sigma_3 = 1$ and recover it thanks to the following rescaling,
\begin{equation}
\rho \rightarrow \rho \, \sigma_3^{2B+A} \,, \qquad \sigma_{1,2} \rightarrow \sigma_{1,2} \, \sigma_{3}^{-1} \ .
\end{equation}
From now on we then only consider the dependence on $\rho, \tau, \sigma_{I=1,2}$.

To illustrate the derivation of the potential, let us consider the 6d Ricci scalar obtained on general grounds from \eqref{Ricci}
\begin{equation}
\label{ricci6}
-2 \Rc_6 = \f{b}{ac} \, \f{a}{be} \, \delta^{ec} + \frac{1}{2} \, \f{a}{ef} \, \f{g}{bc} \, \delta^{eb} \,\delta^{fc} \, \delta_{ga} \,.
\end{equation}
Around their background value, the metric fluxes are expressed as
\begin{equation}
\label{fluctf}
\f{a}{bc} = (\f{a}{bc})^0 \, \rho^{-\frac{1}{2}} \prod_{I} \sigma_I^{\frac{1}{2}(P_I(a)-P_I(b)-P_I(c))} \ ,
\end{equation}
where $P_I(a) = A$ if $a \in \{\api \}$, and $P_I(a) = B$ if $a \in \{\abi \}$. For instance, one has
\begin{equation}
\f{1}{35} = (\f{1}{35})^0 \, \rho^{-\frac{1}{2}} \, \sun^{\frac{A}{2}-B} \, \sde^{-\frac{A}{2}} \ .
\end{equation}
From \eqref{fluctf}, we see that the 6d Ricci scalar \eqref{ricci6} gets an overall factor of $\rho^{-1}$ from its fluctuation along the volume modulus. We then focus on its fluctuations along $\sun, \sde$, denoted $\Rc_6 (\sun,\sde)$ for simplicity. It splits into 6 distinct pieces, as follows
\bea
\Rc_6 (\rho, \sun,\sde) =\, & \rho^{-1} \Rc_6 (\sun,\sde) \label{R6sigma}\\
=\, & \rho^{-1}\left( R_1 \, \sun^{-8} \sde^{4} +
R_2 \, \sun^{4} \sde^{-8} +
R_3 \, \sun^{4} \sde^{4} +
R_4 \, \sun^{-2} \sde^{-2} +
R_5 \, \sun^{4} \sde^{-2} +
R_6 \, \sun^{-2} \sde^{4} \right) \ ,\nn\\
{\rm with}\ & R_1 = - \frac{1}{2}\sum \left(\f{a_{||_1}}{a_{||_2}a_{||_3}} \right)^2 \,, \qquad
R_2 = - \frac{1}{2}\sum \left(\f{a_{||_2}}{a_{||_3}a_{||_1}} \right)^2 \ , \nn\\
& R_3 = - \frac{1}{2}\sum \left(\f{a_{||_3}}{a_{||_1}a_{||_2}} \right)^2 \,, \qquad
R_4 = - \sum \f{a_{||_1}}{a_{||_2}a_{||_3}} \f{a_{||_2}}{a_{||_1}a_{||_3}} \ , \nn\\
& R_5 = - \sum \f{a_{||_2}}{a_{||_3}a_{||_1}} \f{a_{||_3}}{a_{||_2}a_{||_1}} \,, \
R_6 = - \sum \f{a_{||_3}}{a_{||_1}a_{||_2}} \f{a_{||_1}}{a_{||_3}a_{||_2}} \ ,\nn
\eea
where we used the $\f{a}{bc} $ entering our variables \eqref{variables}. In the above, each sum contains 8 terms, and we dropped the label $^0$ for readability. We give these terms explicitly in \eqref{coeff}.

For the fluxes and sources contributions to the potential, we follow \cite{Andriot:2019wrs}. For the fluxes in particular, the powers of $\sigma_I$ are determined by the number $n$ of legs parallel to a given set of sources. For RR fluxes, there is only one $n$ for all sets, so the $\sigma_I$ enter with the same power. For $H$, the two components $H^{(0)}, H^{(2)}$ get exchanged under the two sets $I=1, 2$ as one can see in our variables \eqref{variables}. We obtain
\bea
& |F_q|^2 =  \rho^{-q} (\sun \sde)^{-n A-(q-n) B} |F_q^0|^2 \quad \text{(with}\ F_q = F_q^{(n)} \text{)} \ ,\\
& |H|^2 = \rho^{-3} \sde^{-2A-B} \sun^{-3B} \, (|H^{(0)_1}|^2)^0 + \rho^{-3} \sun^{-2A-B} \sde^{-3B} \, (|H^{(2)_1}|^2)^0 \ .
\eea
We eventually obtain the following 4-field scalar potential (using the simplified notation where we drop integrals and the label ${}^0$)
\bea
\frac{2}{M_p^{2}} \, V(\rho, \tau, \sun, \sde) = &-\tau^{-2}  \rho^{-1} \Rc_6(\sun, \sde) \label{potential}\\
& +\frac{1}{2}\, \tau^{-2} \rho^{-3} \left( \sde^{-2A-B} \sun^{-3B} \, |H^{(0)_1}|^2 + \sun^{-2A-B} \sde^{-3B} \, |H^{(2)_1}|^2 \right) \nn\\
&- \, g_s \, \tau^{-3} \, \rho^{-\frac{1}{2}} \, \left(\sun^A  \sde^B \, \frac{T_{10}^1}{6}+\sun^B \sde^A \, \frac{T_{10}^2}{6}+ \sun^B  \sde^B \, \frac{T_{10}^3}{6}\right) \nn\\
&+ \frac{1}{2}g_s^2\, \tau^{-4} \left(\rho^{2} (\sun \sde)^{-B} |F_1|^2 + (\sun \sde)^{-A-2B} |F_3|^2 \right) \ ,\nn
\eea
where we set from now on $F_5=0$ for simplicity (see \cite{Andriot:2020lea} for the derivation of this term); this flux vanishes in all our solutions. Setting to their background value $\sigma_1=\sigma_2=1$, one recovers the standard 2-field potential $V(\rho, \tau)$, first derived in IIA in \cite{Hertzberg:2007wc}. We will also make use of it.

We can now compute the coefficients for each of our 17 de Sitter solutions. For instance, for solution 16 given as an example in \eqref{solex}, the potential reads
\bea
\frac{2}{M_p^{2}} \, V(\rho, \tau, \sun, \sde) & = \, \tau^{-2} \rho^{-1} \big(0.054981 \, \sun^{4} \sde^{-8} + 0.082159 \, \sun^{-2} \sde^{-2} \label{potential_example} \\
& \phantom{ = \, \tau^{-2} \rho^{-1} } + 0.76145 \, \sde^4 \sun^{-8} + 0.0010436 \, \sun^4 \sde^4\big) \nn\\
&+ \frac{1}{2} \, \tau^{-2} \rho^{-3} \left(0.010366 \, \sun^6 \sde^{-6} + 0.086538 \, \sde^6 \sun^{-6}\right) \nn\\
&- \frac{1}{6} \, \tau^{-3} \rho^{-\frac{1}{2}} \left(1.0654 \, \sun^2 \sde^{-4} + 10 \, \sde^2 \sun^{-4} -0.28655 \, \sun^2 \sde^2\right) \nn\\
&+\frac{1}{2} \, \tau^{-4} \, \left(1.5999 + 0.14675 \, \rho^2 \sun^{-2} \sde^{-2}\right) \ .\nn
\eea
We check, for each of our 17 solutions, that the first derivatives $\del_{\phi^i} V$ all vanish at $\rho = \tau = \sun = \sde = 1 $, and that we precisely recover the value $\Rc_4 = \frac{4}{M_p^{2}} V(1,1,1,1)$. These consistency checks between 10d and 4d were shown formally in \cite{Andriot:2019wrs}.

For completeness, let us consider additional contributions $\bar{T}_{10}^I$ from $\bar{D}_5$, mentioned at the end of section \ref{sec:eq}. From the derivation of the potential, one can see that only the DBI term of the source action contributes, and this term has the same sign as for $D_5$. Therefore, in the potential, one should simply make the replacement
\beq
T_{10}^I \rightarrow T_{10}^I + \bar{T}_{10}^I \ .
\eeq
We now do not consider $\bar{D}_5$ any further, but we will come back to them in section \ref{sec:ccl}.\\

The above potential is the one entering \eqref{S4dgen}. We now rewrite this 4d action with its kinetic terms, computed on general grounds in appendix \ref{ap:4d}
\bea
{\cal S} = \int \d^4 x \sqrt{|g_4|} &\, \Bigg( \frac{M_p^2}{2} \Rc_4 - V(\rho, \tau, \sun, \sde) \label{S4dkin1}\\
&\, - \frac{M_p^2}{2} \bigg(\frac{3}{2\rho^2} (\del \rho)^2 + \frac{2}{\tau^2}(\del \tau)^2  + 12 \Big(\frac{1}{\sun^2}(\del \sun)^2  + \frac{1}{\sde^2}(\del \sde)^2 - \frac{1}{\sun\sde}\del_{\mu}\sun \del^{\mu}\sde  \Big)\bigg) \Bigg) \nn\\
 = \int \d^4 x \sqrt{|g_4|}  &\, \Bigg( \frac{M_p^2}{2} \Rc_4 - V(\rho, \tau, \sun, \sde) \label{S4dkin2}\\
&\, - \frac{M_p^2}{2} \Big(\frac{3}{2} (\del \ln \rho)^2 + 2 (\del \ln \tau)^2  + 9 \big(\del \ln \frac{\sun}{\sde}\big)^2  + 3 \big(\del \ln (\sun \sde )\big)^2 \Big)  \Bigg) \ .\nn
\eea
From \eqref{S4dkin2} we can read-off canonically normalized scalar fields. Since the potential is written in terms of $\sigma_{1,2}$, we will rather use \eqref{S4dkin1} at the cost of having a non-diagonal field space metric $g_{ij}$. It is given by
\begin{equation}
g_{ij} = M_p^2\,
\begin{pmatrix}
\mathlarger{\frac{3}{2 \rho^2}} & 0 & 0 &0\\[10pt]
0 & \mathlarger{\frac{2}{\tau^2}} & 0 &0 \\[10pt]
0 & 0 & \mathlarger{\frac{12}{\sun^2}} &\mathlarger{-\frac{6}{\sun \sde}} \\[10pt]
0 & 0 &\mathlarger{-\frac{6}{\sun \sde}} & \mathlarger{\frac{12}{\sde^2}}
\end{pmatrix} \ . \label{gij}
\end{equation}
We now have all the tools to study the stability of our de Sitter solutions.

\subsection{Stability analysis}\label{sec:stabana}

Each of our 17 de Sitter solutions matches in 4d an extremum of the potential $V$ \eqref{potential} along the four fields $\phi^i \in \{\rho, \tau, \sun, \sde \}$, at $\phi^{i=1,..,4}=1$. To study the stability of such a solution, we consider the mass matrix $g^{ik} \nabla_k \del_j V= g^{ik} H_{kj}$, with $\del_i \equiv \del / \del \phi^i$ and the field space metric $g_{ij}$ given in \eqref{gij}. At an extremum, the Hessian $H_{jk}$ is computed as follows
\begin{equation}
\label{hessian}
H_{jk} \equiv \nabla_j \del_k V = \del_j \del_k V - \Gamma^i_{jk} \cancelto{0}{\del_i V} =  \del_j \del_k V \ .
\end{equation}
We then compute the mass matrix and its eigenvalues, which correspond to $\text{masses}^2$. If one eigenvalue is negative, we have a tachyon. Note that since $g^{ik}$ is positive definite, the eigenvalues of the Hessian have the same signs, so one can also read off the presence of a tachyon from $H_{jk}$. By definition, the latter corresponds to a maximum of the potential in one direction. We further obtain the eigenvector associated to the negative eigenvalue of the mass matrix, to deduce the tachyonic direction in field space (at the extremum): we denote it $\vec{v}$, specified along $(\rho, \tau, \sun, \sde)$.

The result is that for each of our 17 de Sitter solutions, there is always one unstable direction, corresponding to a tachyon. For 11 solutions, the tachyon is found already within the $(\rho,\tau)$-subspace, meaning that studying the 2-field potential $V(\rho,\tau)$ is sufficient (see Table \ref{tab:eta}). For the 6 other solutions, one needs the four fields. This is consistent with the proposal of \cite{Danielsson:2012et}, that the tachyon should systematically lie in this 4-field space, and sometimes already among the two fields $(\rho,\tau)$.

\begin{figure}[H]
\begin{center}
\includegraphics[width=0.7\textwidth]{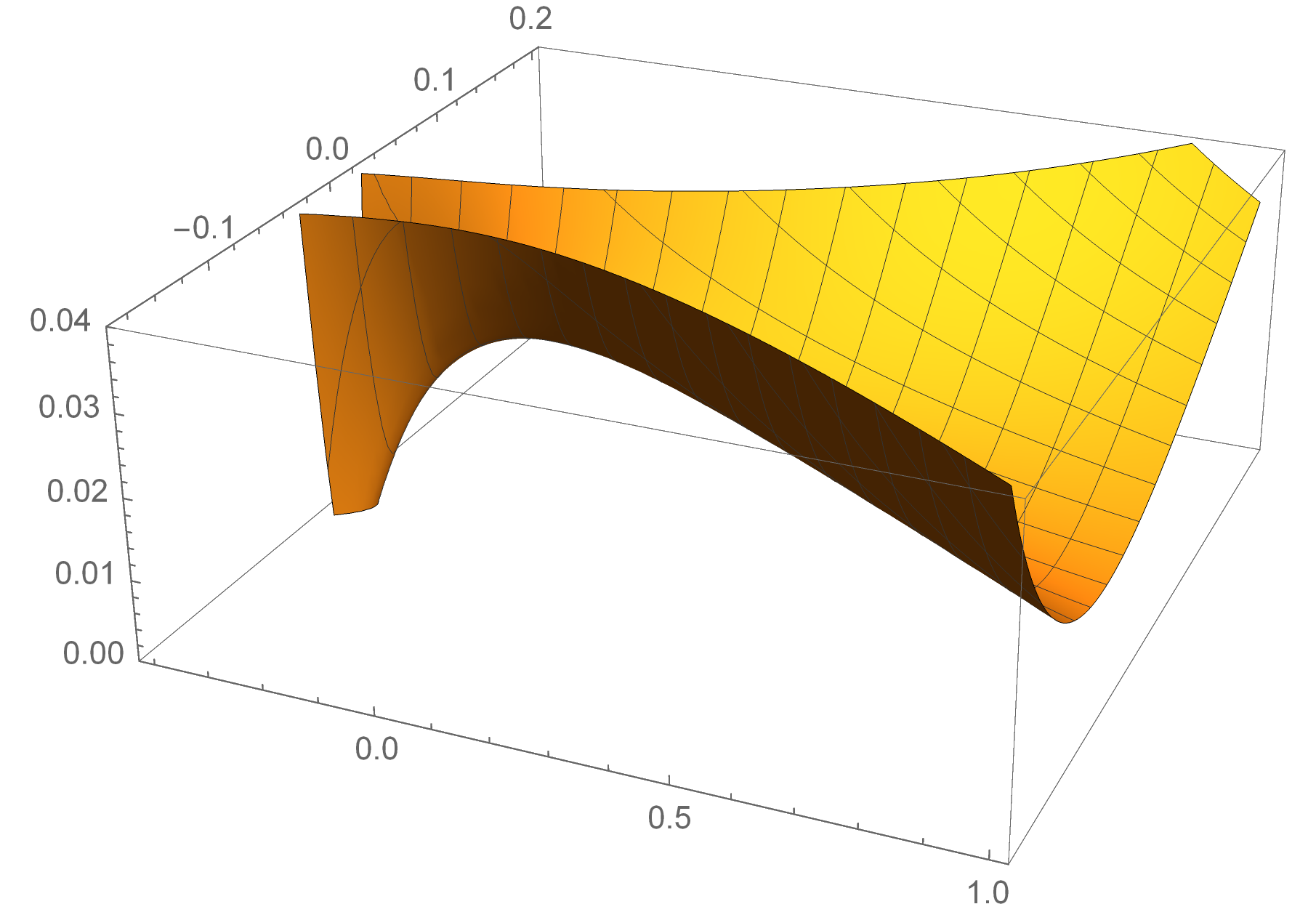}
\caption{Potential $\frac{2}{M_p^2} V(\rho, \tau, \sun, \sde)$ around its de Sitter extremum corresponding to solution 16. The potential is displayed along a fluctuation $s$ around $\sun=1$, and another one $t$ along the tachyonic direction $(1,1,1,1) + t \vec{v}$, where $\vec{v}$ is the eigenvector associated to the negative eigenvalue of the mass matrix, at the extremum. At $s=t=0$, we verify the positive maximum along one direction, and minimum along the other one.}\label{fig:celineVtachyon}
\end{center}
\end{figure}

As an example, for solution 16 given in \eqref{solex}, the eigenvalues of the mass matrix\footnote{One verifies that $M_p$ drops out of the mass matrix. The dimension of the masses is then given by that of the coefficients in the scalar potential. As we will see in section \ref{sec:regime}, the unit for a mass is then $1/(2\pi l_s)$.} and the tachyonic eigenvector are given by
\begin{align}
\text{masses}^2 &= (1.6235,\, 0.26174,\, 0.12567,\, -0.035395) \ , \\ \vec{v} &= (0.48957,\, 0.83657,\, 0.20509,\, 0.13567) \ .\nn
\label{eigen16}
\end{align}
We observe that the contribution of each field direction to $\vec{v}$ is similar for every solution (see appendix \ref{ap:sol}), with in particular most of the tachyon carried by the $\tau$ direction. We illustrate this instability in Figure \ref{fig:celineVtachyon} by displaying the 4-field potential along the tachyonic direction at the extremum, and along another, stabilized, direction.

To characterize more precisely the observed instability, we compute the parameter $\eta_V$
\beq
\eta_V = \frac{M_p^2}{V} \times \text{minimal eigenvalue of } g^{ij} \nabla_j \del_k V \ , \label{eta}
\eeq
at the extremum, where $\phi^i=1$. The values of $\eta_V$ obtained for the 17 solutions are listed in Table \ref{tab:eta}, both for the 2-field and 4-field potential. For the latter, we find that $\eta_V \in \left[-2.9703, -1.7067\right]$, with an average value of $-2.7000$ and a median value of $-2.8544$. This is in good agreement the refined de Sitter swampland conjecture of \cite{Ooguri:2018wrx}: the latter requires $\eta_V \leq -c' \sim -\mathcal{O}(1)$.
\begin{table}[h]
  \begin{center}
  		\hspace*{-0.5cm}
    \begin{tabular}{|c||c|c|c|c|c|c|c|c|c|}
    \hline
\mbox{Solution}  & 1 & 2 & 3 & 4 & 5 & 6 & 7 & 8 & 9 \\
    \hhline{==========}
$-\eta_V\ \text{2-field} $ & 0.46553
 & 1.0463
  & 1.1819
   & 1.0778
    & 1.2315
     & 0.95707
      & 1.0209
       & 0.57612
        & 0.96718

            \\
    \hhline{-||---------}
$-\eta_V\ \text{4-field}$ & 2.8544
 & 2.7030
  &
2.9334
 &
2.8966
 &
2.9703
 &
2.9146
 &
2.5101
 &
2.7790
 &
2.2494

  \\
    \hline
  \multicolumn{10}{c}{}\\
    \hhline{---------~}
\mbox{Solution}  & 10 & 11 & 12 & 13 & 14 & 15 & 16 & 17 & \multicolumn{1}{c}{} \\
    \hhline{=========~}
$-\eta_V\ \text{2-field}$ & $\cdot$ & 1.1916
 & $\cdot$ & $\cdot$ & $\cdot$ & 0.17914
  & $\cdot$ & $\cdot$ & \multicolumn{1}{c}{} \\
    \hhline{-||--------~}
$-\eta_V\ \text{4-field}$ &2.0908

&2.9354

&2.7548

&2.9518

&1.7067

&2.9336

&2.8404

&2.8748
 & \multicolumn{1}{c}{}
 \\
    \hhline{---------~}
    \end{tabular}
     \caption{Values of $-\eta_V$ obtained for each of our 17 de Sitter solutions, either with the 2-field potential $V(\rho,\tau)$ or the 4-field one $V(\rho,\tau,\sun,\sde)$. A dot $\cdot$ means that $\eta_V (\rho,\tau) > 0$.}\label{tab:eta}
  \end{center}
\end{table}

We can compare these values of $\eta_V$ to those obtained for type IIA solutions (see e.g.~\cite{Andriot:2018mav} and \cite{Caviezel:2008tf, Flauger:2008ad}): most of the latter are below $-3.6$, with the exception of one at $-2.5$. Similarly, for the only solution found in IIB so far \cite{Caviezel:2009tu}, the value obtained is $-3.1$. To compute these older values of $\eta_V$, all scalar fields of a 4d effective action were used, while here we only considered a subset of 2 or 4 fields. For this reason, the values in Table \ref{tab:eta} should actually be considered as an upper bound on $\eta_V$, and this may explain the small difference with the older values. We discuss this point in greater detail in the next subsection.

\subsection{A mathematical property of the mass matrix}\label{sec:lemma}

Let us introduce the following relevant lemma, reminiscent of the Sylvester criterion (see \cite{Shiu:2011zt}).

\begin{lemma}
Let $M$ be a square symmetric matrix of finite size, and $A$ an upper left square block of $M$. Let $\mu_1$ be the minimal eigenvalue of $M$ and $\alpha$ any eigenvalue of $A$. Then one has $\mu_1 \leq \alpha$.
\end{lemma}

\begin{proof}
Let us denote by $n$ the size $n\times n$ of $M$. Because $M$ is symmetric, it can be diagonalised to $D$ with orthogonal matrices $O$, such that $M =O D O^\top $.  Let $\mu_{i=1,\dots,n}$ be the eigenvalues of $M$, i.e.~the diagonal entries of $D$. Up to relabelling, we call $\mu_1$ the smallest, i.e. $\mu_1 \leq \mu_i$. Let us now consider any $n$-vector $X$ of components $x^i$ and $X'=O^\top X$ of components ${x'}^i$. One has $\sum_i (x^i)^2=||X||^2=||X'||^2$. We further introduce the quadratic form
\beq
q(X)=X^\top M X = {X'}^\top D X'= \sum_i \mu_i ({x'}^i)^2 \geq \sum_i \mu_1 ({x'}^i)^2 = \mu_1 ||X||^2 \ . \label{proofineq}
\eeq
We now turn to $A$ which is also symmetric. We then consider an eigenvector $Y$ of $A$ (thus of non-zero norm), with eigenvalue $\alpha$: $AY = \alpha Y$. We complete $Y$ to a $n$-vector $\tilde{Y}$ by adding $0$'s as components. We then compute $q(\tilde{Y}) =Y^\top A Y = \alpha  ||Y||^2 =  \alpha  ||\tilde{Y}||^2$. Using the inequality \eqref{proofineq}, we deduce $\mu_1 \leq \alpha$.
\end{proof}
This lemma has important consequences regarding the mass matrix and $\eta_V$. From it, we conclude that adding more fields to a theory, i.e.~adding lines and columns to the mass matrix, can only lower its minimal eigenvalue. It implies that~$\eta_V$ will be lowered when including more fields. It also implies that a tachyon cannot be removed by introducing more fields, rather the lowest tachyonic squared mass would only get lowered. This is conceptually important since we consider most of the time theories that are truncations to a finite set of fields.

As mentioned above, this could explain why our values for $\eta_V$ are higher than those obtained in previously known solutions. It is also fair to say that the settings are different: previous solutions were in type IIA, or type IIB with different sources. There is thus a priori no reason to recover the same values. Another illustration of this idea is the comparison between the $\eta_V$ value computed with the 2-field potential, and that with the 4-field one. For the 2-field cases having a tachyon, we obtain $\eta_V \in \left[-1.2315, -0.17914\right]$, with an average value of $-0.89956$ and median value of $-1.0209$. This is significantly higher than the $\eta_V$ obtained with all four fields, as can be seen in Table \ref{tab:eta}. Once again, $\eta_V$ is lowered with more fields.\\

\begin{figure}[h]
\begin{center}
  \begin{subfigure}[H]{0.4\textwidth}
	\includegraphics[width=\textwidth]{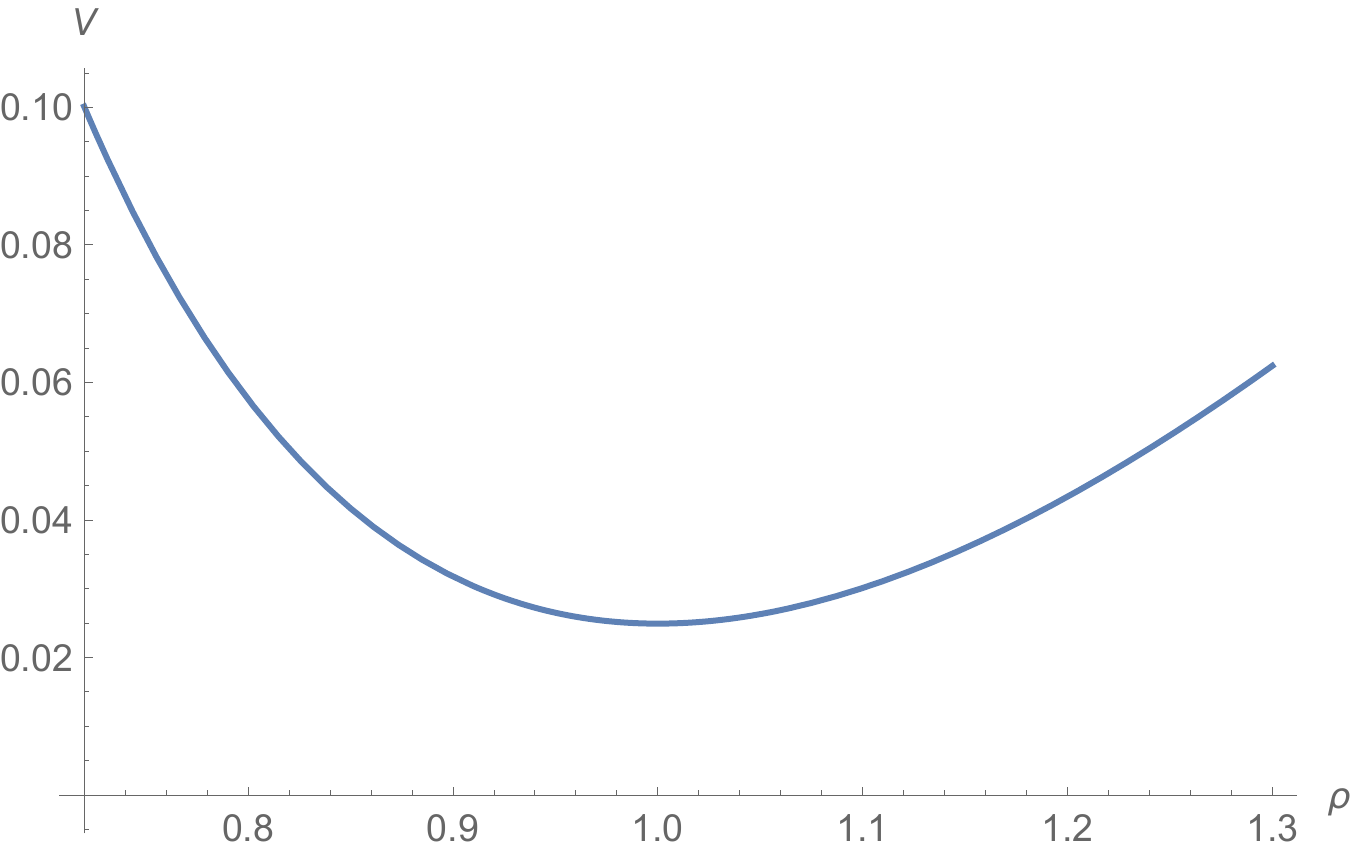}
	\caption{$\ \ (\rho,1,1,1)$}\label{fig:Vrho}
\end{subfigure}
\qquad \qquad
  \begin{subfigure}[H]{0.4\textwidth}
	\includegraphics[width=\textwidth]{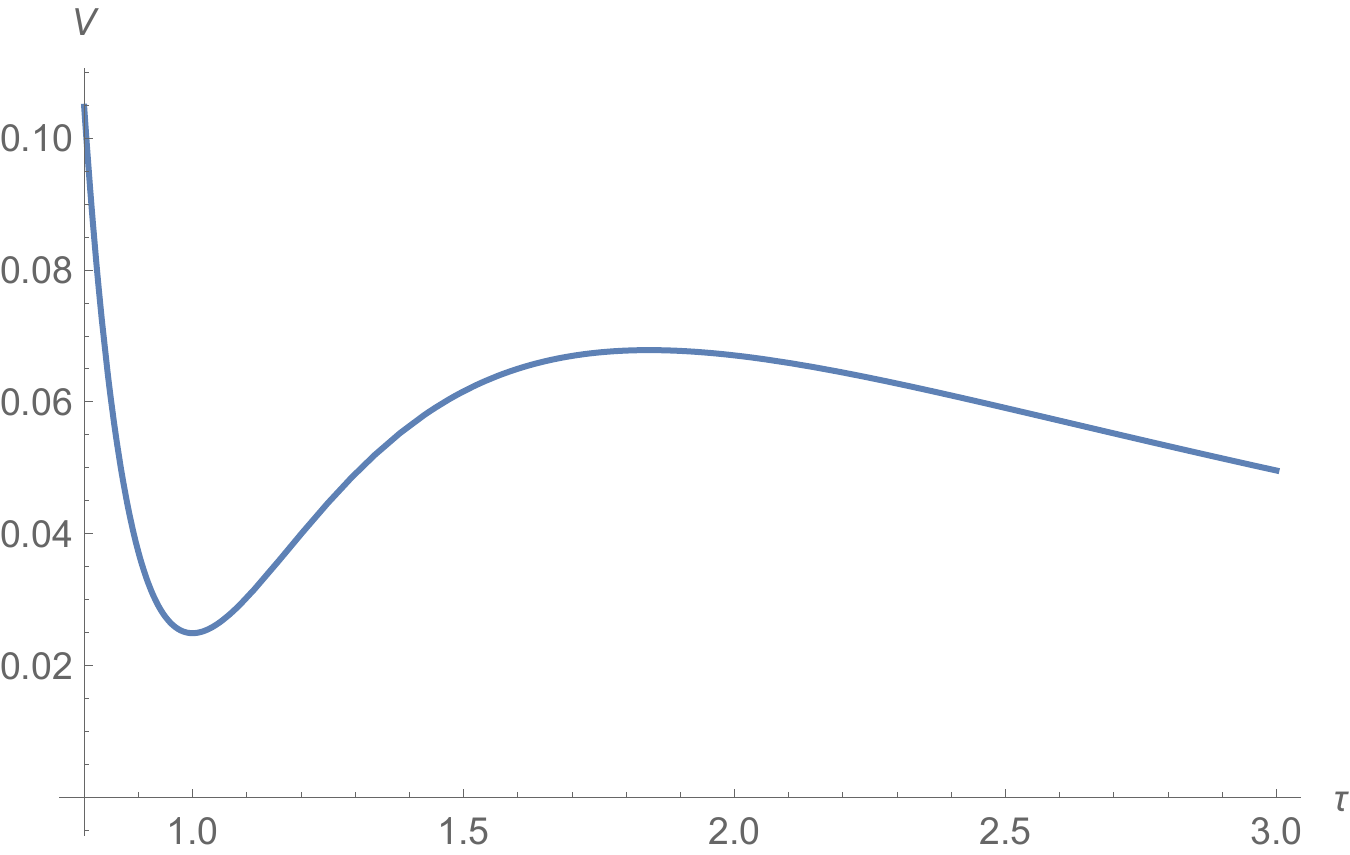}
	\caption{$\ \ (1,\tau,1,1)$}\label{fig:Vtau}
\end{subfigure}
  \begin{subfigure}[H]{0.4\textwidth}
	\includegraphics[width=\textwidth]{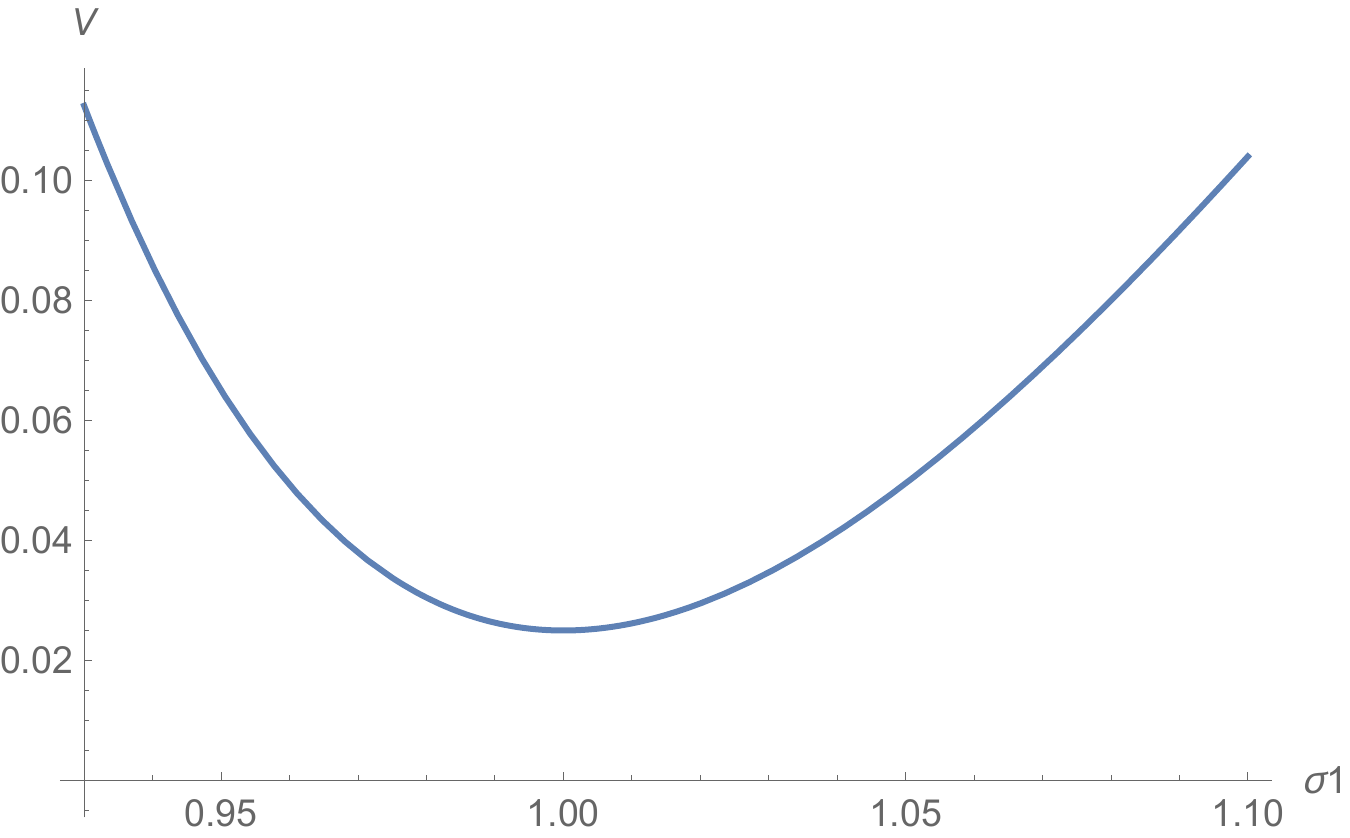}
	\caption{$\ \ (1,1,\sun,1)$}
\end{subfigure}
\qquad \qquad
  \begin{subfigure}[H]{0.4\textwidth}
	\includegraphics[width=\textwidth]{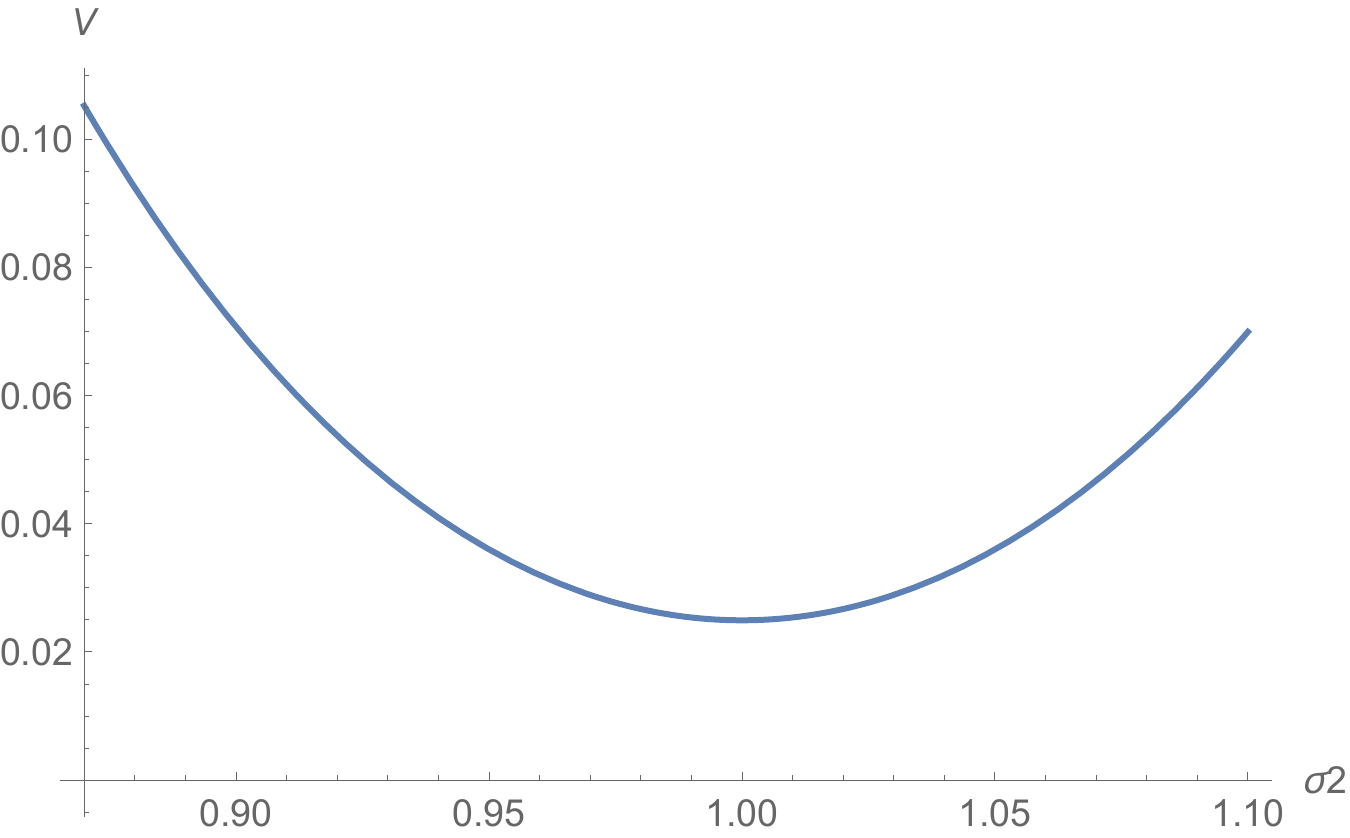}
	\caption{$\ \ (1,1,1,\sde)$}
\end{subfigure}
\caption{Slices of the potential $\frac{2}{M_p^2} V(\rho, \tau, \sun, \sde)$ along each of the four fields, close to the de Sitter extremum at $\phi^i=1$.}\label{fig:potential1}
\end{center}
\end{figure}

Finally, let us consider the potential along each of the fields $(\rho, \tau, \sun, \sde)$ close to the de Sitter solution. For instance, we display it in Figure \ref{fig:potential1} for solution 16 given in \eqref{solex}.
For each of our solutions, we observe that $\del_{\phi^i}^2 \, V > 0$ at the extremum, for each of the four fields. We verify that every diagonal entry of the mass matrix at the extremum is also positive, for each solution. This is reminiscent of the results of section 3.3 of \cite{Andriot:2018ept} on stability (see however a note on those in \cite{Andriot:2020lea}). It was shown there that $\del_{\rho}^2 \, V > 0$ for such classical de Sitter solutions, irrespectively of whether sources are parallel or intersecting. We deduce from this observation that the tachyon, at least in our solutions, always stems from the off-diagonal entries of the Hessian or of the mass matrix. This is consistent with the above lemma. If true in general, this observation indicates that using only diagonal entries to prove the presence of a tachyon is then not appropriate. Such a method was developed in \cite{Andriot:2018ept, Andriot:2019wrs}, and the present observation may explain why this attempt failed.

This discussion illustrates also the fact that one should be careful with apparent extrema of the scalar potential that appear in slices of field space. For instance, we noticed for all our solutions that the potential displayed along the two fields $(\rho, \tau)$ or some other subset of the 4-field directions $(\rho,\tau, \sigma_1, \sigma_2)$ exhibited, away from our 10d solution, other apparent critical points. Some of these even looked like de Sitter minima. However, such points, which appeared critical along some field directions, were not found to be extrema in all four field directions. Finding a de Sitter solution thus remains a non-trivial result.

\section{The classical regime of string theory}\label{sec:regime}

The main motivation for having de Sitter solutions of 10d type II supergravities is that they may correspond to classical and perturbative backgrounds of string theory. This would hold if some conditions are satisfied. For instance, one usually requires a small $g_s$ and a large internal 6d volume, to neglect string loop and $\alpha'$ corrections; we detail these requirements more precisely in section \ref{sec:settingquantiz}. Recent works \cite{Roupec:2018mbn, Junghans:2018gdb, Banlaki:2018ayh, Andriot:2019wrs, Grimm:2019ixq} have argued, in various settings, in favor of an absence of de Sitter solutions in asymptotic regimes of string theory, e.g.~in a parametrically controlled classical perturbative regime corresponding to 10d supergravity. This is in line with swampland conjectures \cite{Ooguri:2018wrx, Bedroya:2019snp}, that generalise the Dine-Seiberg argument \cite{Dine:1985he}. In other words, as summarized in conjecture 3 of \cite{Andriot:2019wrs}, even though one may find de Sitter solutions of 10d supergravities, those may not satisfy the conditions that would establish them as classical string backgrounds. Loopholes to such arguments have however been pointed-out in \cite{Hebecker:2018vxz, Junghans:2018gdb, Andriot:2019wrs}. In particular, an important distinction to be made is the difference between some scalar fields having large but finite values, versus the asymptotic behaviour and infinite distance limit in field space. The former may actually be more relevant for physics: an obvious example is the size of the internal dimensions that has to be much larger than the string length, but small when compared to observational bounds. Such a ``grey zone'' in field space could accommodate interesting classical de Sitter solutions, that may not survive in an asymptotic limit.

The aim of this section is to test our de Sitter solutions in that respect. We introduce the necessary tools in section \ref{sec:settingquantiz} and present more precisely the conditions to be verified. We use a 10d language as in \cite{Andriot:2019wrs}, rather than performing a more common 4d study of the volume and dilaton; we will see that this allows more precise checks, and discuss the relation between the two. We start testing the validity of our solutions in the classical regime in section \ref{sec:firstcheckquantiz}. A certain subset of the requirements are successfully checked on 4 solutions, going beyond what has been done previously in the literature and shedding light on the above discussion. However, a complete check is only possible with a detailed knowledge of the 6d geometry. Indeed, a proper flux quantization, the count of the number of orientifolds, and the lattice quantization conditions (involving the $f^a{}_{bc}$) all require to know precisely the 6d group manifold. As explained in section \ref{sec:compactness}, it is only the case for solutions 14 and 15. Their complete study is delayed to a companion paper \cite{Andriot:2020vlg}.

\subsection{Requirements and setting}\label{sec:settingquantiz}

The first requirement for a classical string background is $g_s \ll 1$, to neglect quantum corrections. In addition, to avoid $\alpha'$ corrections, all lengths should be bigger than the string length $l_s$. If our 6d manifold $\mmm$ is made of 6 circles of radius $r^{a=1,\dots,6}$, we should require $r^a \gg l_s$. Two stringy requirements have in addition to be fulfilled, for the supergravity quantities to match string objects. First, the number of orientifolds $N_{O_p}^I$ in each set $I$ is determined by the geometry: it corresponds to the number of fixed points under the involution in the transverse directions, and is thus finite. Secondly, the harmonic components of fluxes, i.e.~the fluxes through cycles in homology, have to be quantized: they are given in terms of integers denoted below $N_{q\, a_1 \dots a_q}$. Finally, a requirement is inherent to having a compact group manifold (see section \ref{sec:compactness}): the lattice imposes quantization conditions on the structure constants $f^a{}_{bc}$. We summarize these five requirements as follows
\beq
g_s \ll 1 \ , \ \ r^a \gg l_s \ ,\ \ N_{O_p}^I \ \mbox{finite}\ ,\ \ N_{q\, a_1 \dots a_q} \in \mathbb{Z} \ , \ \ f^a{}_{bc} \ \mbox{quantized} \ .\label{requir}
\eeq

Several of these requirements need a detailed knowledge of the group manifold geometry, namely an explicit expression of globally defined $e^a$, with the lattice action on local coordinates. It is obvious for the quantization of $f^a{}_{bc}$, to start with. Determining the fixed points of the orientifold involution, and thus the value of $N_{O_p}^I$, also depends on this knowledge. Finally, determining the harmonic components of fluxes depends in practice on knowing explicitly the cycles and corresponding $e^a$; the condition on $N_{q\, a_1 \dots a_q}$ is then affected. The mere values of the structure constants $f^a{}_{bc}$, known for each solution, is not straightforwardly giving us the knowledge of the geometry, as explained in section \ref{sec:compactness} and appendix \ref{ap:alg}. In particular, changes of basis are typically needed to reach appropriate $e^a$. For this reason, we will only perform in the following some checks among the list \eqref{requir}, giving a first illustration of such an analysis.\\

Let us define the quantities entering the requirements \eqref{requir}, and relate them to the variables in our 10d solutions. We follow notations of section 5 of \cite{Andriot:2019wrs}. We first introduce ``radii'' formally as follows
\beq
2\pi r^a = \int e^a \ . \label{oneformint}
\eeq
These one-form integrals are strictly speaking not necessarily well-defined, and this should rather be viewed as a normalization convention. These integrals depend on the details of the geometry, and we can thus not compute them generically. In practice, we will only need certain combinations of one-forms entering e.g.~appropriate flux components. So we only assume for now that the relevant one-form combinations provide well-defined integrals, compatible with the convention \eqref{oneformint}. The flux quantization condition along directions $e^{a_1} \w \dots \w e^{a_q}$ is then written as
\beq
\frac{1}{(2\pi l_s)^{q-1}} \int_{a_1 \dots a_q} F_q = N_{q\, a_1 \dots a_q} \in \mathbb{Z} \ \Rightarrow \ F_{q\, a_1 \dots a_q}= \frac{N_{q\, a_1 \dots a_q}}{2\pi l_s} \frac{l_s^q}{r^{a_1} \dots r^{a_q}} \ ,
\eeq
valid also for the $H$-flux. In absence of more knowledge on the geometry, we ask for a quantization of all flux components, which is actually overconstraining since some fluxes might not be in cohomology. We also trade the integer condition for simply having numbers bigger than $1$: indeed, since a change of basis could be needed, having precise integers does not make sense at this stage.

We turn to the sources. Using the definition of $T_{10}^I$ \cite{Andriot:2016xvq} and the smeared ansatz, we obtain the general expression
\beq
\frac{T_{10}^I}{p+1} = (2^{p-5} N_{O_p}^I - N_{D_p}^I) \frac{(2\pi l_s)^{7-p}}{\sqrt{g_{\bot_I}}} \ \Rightarrow\ {\rm Here:}\ \frac{T_{10}^I}{6} = \frac{N_s^I}{(2\pi l_s)^2} \frac{l_s^{4}}{r^{a_{1\bot_I}} \dots r^{a_{4\bot_I}}} \ , \label{T10I}
\eeq
with the number of sources given by $N_s^I=N_{O_5}^I - N_{D_5}^I$. The fixed value of $N_{O_5}^I$ thus gives an upper bound to $N_s^I$. On our group manifolds, this number is likely to be comparable to that on a torus, which is $N_{O_5}^I = 16$ for an $O_5$. In the following, we first restrict to $N_s^I \leq 100$, before turning to $16$. Here again, we trade the integer condition for numbers bigger than 1 (in absolute value).

Finally, even if not addressed here, let us mention the structure constants. The scaling of $f^a{}_{bc}$ with the radii is clear from the Maurer-Cartan equations, so we can introduce the numbers $N_{abc}$ as follows
\beq
f^a{}_{bc} = \frac{r^a}{r^b r^c} \frac{N_{abc}}{2\pi} = \frac{N_{abc}}{2\pi l_s}\, \frac{r^a l_s}{r^b r^c} \ .
\eeq
The quantization of the structure constants depends on the lattice of the compact manifold. It translates, in each case, into some discretization conditions on the $N_{abc}$.\\

From these definitions, we see that the dimension of these quantities is given by powers of the fundamental length $2\pi l_s$. We can consistently consider adimensional quantities, by multiplying them by the appropriate power of $2\pi l_s$. Equivalently, one verifies that all e.o.m.~and BI are invariant under the following scaling
\beq
F_{q\,{\rm new}} = 2\pi l_s\, F_{q\,{\rm old}} \ , \ f^a{}_{bc\,{\rm new}} = 2\pi l_s\, f^a{}_{bc\,{\rm old}} \ , \ T_{10\,{\rm new}}^I = (2\pi l_s)^2\, T_{10\,{\rm old}}^I \ .\label{adim}
\eeq
The new adimensional quantities, with appropriate $g_s$ factors as in \eqref{bla}, are the ones to be identified with the numerical values obtained in our solutions. Furthermore, we express from now on the radii in units of $l_s$, i.e.~redefine them as
\beq
r^a_{{\rm new}} = \frac{r^a_{{\rm old}}}{l_s} \ ,
\eeq
such that the physical requirement becomes $r^a \gg 1$. We now summarize the problem to be solved: given our de Sitter solutions, one should find a set of variables $\{g_s, r^a, N_{q\, a_1 \dots a_q}, N_s^I \}$, defined as follows
\beq
g_s F_{q\, a_1 \dots a_q}= \frac{g_s\, N_{q\, a_1 \dots a_q}}{r^{a_1} \dots r^{a_q}} \ , \quad g_s T_{10}^I = \frac{6 g_s \, N_s^I}{r^{a_{1\bot_I}} \dots r^{a_{4\bot_I}}} \ ,\label{bla}
\eeq
such that
\beq
0 < g_s \leq 10^{-1} \ , \ r^a \geq 10 \ , \ N_s^I \leq 100 \ ,\ |N_s^I| \geq 1 \ ,\  |N_{q\, a_1 \dots a_q}| \geq 1  \ . \label{constraintsNs}
\eeq
Having a hierarchy of order 10 is a minimal requirement, to test the possibility of reaching a classical regime. Physically, stronger constraints would be preferred, in particular on the radii. At this stage, it is unclear how much room we have for this, but we will come back to it in \cite{Andriot:2020vlg}. We now check the requirements \eqref{constraintsNs} on our solutions.

\subsection{Testing the solutions}\label{sec:firstcheckquantiz}

\subsubsection*{Overall $\lambda$-rescaling}

To test our solutions against the constraints \eqref{constraintsNs}, we consider as a warmup the quantity $g_s T^1_{10}$. In many of our solutions (see appendix \ref{ap:sol}), this quantity was set equal to 10, providing us with a scale when looking for solutions as explained in appendix \ref{ap:num}. We now obtain the following equation to solve
\beq
10 = g_s T^1_{10} = 6 \times g_s \, \frac{1}{r^3 r^4 r^5 r^6} \, N_s^1 \ .
\eeq
Asking for $g_s \leq 10^{-1}$, $r^a \geq 10$, we get $N_s^1 > 10^5$, which is too big: we cannot satisfy the requirements \eqref{constraintsNs}.

One could however consider different solutions, obtained by transforming ours using a symmetry of the e.o.m.~and BI. A typical example of such a transformation is a scaling symmetry. The first one to consider is an overall rescaling, which diminishes every quantity in the solution. It corresponds to an overall scaling of the potential, which does not change its critical points. So we consider a solution tilde, obtained by a scaling with a real parameter $\lambda > 0$
\beq
\tilde{F}_q= \frac{1}{\lambda} F_q \ , \ \tilde{f}^a{}_{bc}= \frac{1}{\lambda} f^a{}_{bc} \ , \ \tilde{T}^I_{10} = \frac{1}{\lambda^2} T_{10}^I \ .
\eeq
This leaves all equations of section \ref{sec:eq} and the constraints \eqref{constraints} invariant, so we obtain again a de Sitter solution. Coming back to our problem, we now satisfy the requirements \eqref{constraintsNs} on the quantity $g_s \tilde{T}^1_{10}$ if we pick $\lambda > 10^2$. From now on, we always include this $\lambda$-rescaling.

\subsubsection*{Equal radii $r^a = r$}

To test our (rescaled) solutions against the constraints \eqref{constraintsNs}, we first consider for simplicity {\it all radii to be equal to one value $r$}. Satisfying the requirements \eqref{constraintsNs} in that case should correspond to the 4d analysis involving only the dilaton and volume. In the literature, several such 4d studies concluded negatively. Here, we verify as well on our 17 solutions that upon this simplification, it is very difficult to satisfy the constraints \eqref{constraintsNs}: it is only possible for 3 solutions, which remain far from having the admissible $N_s^I \leq 16$.

To show this, a simple test can be performed. As identified in \cite{Andriot:2019wrs}, the two following ratios are critical in these discussions
\beq
\frac{g_s^2 F_{1\, a}^2}{g_s T_{10}^1}  \ ,\quad  \frac{H_{a_1 a_2 a_3}^2}{g_s T_{10}^1}  \ ,
\eeq
where we take the smallest of the $F_1$ and $H$ flux components. Interestingly, these ratios are independent of the scaling $\lambda$. With all radii equal to $r$, one obtains
\beq
N_{1\, a}^2 = \frac{g_s^2 F_{1\, a}^2}{g_s T_{10}^1}\ \frac{6 N_s^1}{g_s r^2} \ , \quad N_{H\, a_1 a_2 a_3}^2 = \frac{H_{a_1 a_2 a_3}^2}{g_s T_{10}^1} \ 6 N_s^1\ g_s r^2 \ ,
\eeq
and we take here $T_{10}^1 >0$. Requiring that both flux integers are greater than $1$, we obtain the following inequalities
\beq
N_s^1 \geq \frac{g_s T_{10}^1}{6 g_s^2 F_{1\, a}^2} g_s r^2 \ ,\  N_s^1 \geq \frac{g_s T_{10}^1}{6 H_{a_1 a_2 a_3}^2} \frac{1}{g_s r^2} \ \Rightarrow \ N_s^1 \geq \frac{g_s T_{10}^1}{6 g_s  |F_{1\, a}\, H_{a_1 a_2 a_3}| } \ .
\eeq
Using the last inequality, it is straightforward to show that all 17 solutions, except 3, must have $N_s^1 > 100$, thus violating the constraints \eqref{constraintsNs}. These constraints can be satisfied for the remaining 3 solutions: we then obtain the following values
\beq
{\rm Solution}\ 10\!:\ N_s^1= 44.923  \,,\; {\rm Solution}\ 12\!:\ N_s^1=92.591  \,,\; {\rm Solution}\ 17\!:\ N_s^1=98.719  \ ,
\eeq
where we display the biggest among $N_s^{I=1,2}$. These values remain far from $N_s^I \leq 16$, so we conclude negatively in this simple analysis.

\subsubsection*{Different radii values $r^a$}

Crucially, the situation changes when {\it considering different radii $r^a$}. Allowing them to take different values, e.g.~having internal hierarchies among them, certainly gives more room to satisfy the constraints \eqref{constraintsNs}. The need for such internal hierarchies was already advocated in \cite{Andriot:2019wrs}. We now obtain that 8 of our 17 (rescaled) solutions verify the requirements \eqref{constraintsNs} with different radii values. In addition, 4 solutions admit $N_s^I \leq 16$, as detailed below. This point is then important, and doing such a refined analysis in 4d may also lead to different conclusions than previously obtained.

To start with, the following solutions satisfy the constraints \eqref{constraintsNs}
\bea
&\hspace{-0.1in} {\rm Solution}\ 3\!:\ N_s^2=92.542  \ ,\ \ {\rm Solution}\ 5\!:\ N_s^2=93.546  \ ,\ \ {\rm Solution}\ 11\!:\ N_s^2=91.922  \ , \\
&\hspace{-0.1in} {\rm Solution}\ 14\!:\ N_s^1=30.269 \ , \label{sol14Ns}
\eea
where we display the highest value of $N_s^{I=1,2}$. The solution 14 is special, as discussed around \eqref{lambda1}: it is the only one with $T_{10}^2 < 0$, leading to the automatically satisfied constraint $N_s^2 < 0 <100$, as is already the case for all solutions with $N_s^3$. Then, 4 other solutions satisfy the constraints \eqref{constraintsNs}, now with $N_s^I \leq 16$. The corresponding parameters are given as follows
\bea
& {\rm Solution}\ 10\ (\mbox{with}\ \lambda= 1222.4)\!:\\
& N_s^1=  7.6063 \ ,\ N_s^2= 7.6573  \ ,\ N_s^3= -1.0986  \ , \ g_s= 0.0079125 \ ,\nn\\
&r^1= 19.707  \ ,\ r^2= 24.897 \ ,\ r^3= 20.599 \ ,\ r^4= 23.661 \ ,\ r^5= 10.487 \ ,\ r^6= 10.557 \ ,\nn\\
& N_{1\, 5}= 1.0843 \ ,\ N_{1\, 6}= 1.0915 \ , \ N_{H\, 125} = 3.9469 \ ,\ N_{H\, 126} = -1.2691 \ , \ N_{H\, 345} = 2.7847 \ , \nn\\
&N_{H\, 346} = -3.0850 \ , \ N_{3\, 135} = -59.411 \ ,\ N_{3\, 136} = -330.36 \ ,\ N_{3\, 145} = 240.55 \ , \nn\\
&N_{3\, 146} = -113.37 \ , \ N_{3\, 235} = -256.59 \ , \ N_{3\, 236} = 23.890 \ ,\ N_{3\, 245} = 182.70 \ ,\ N_{3\, 246} = -183.91  \ .\nn
\eea
\bea
& {\rm Solution}\ 12\ (\mbox{with}\ \lambda= 3252.5)\!:\\
& N_s^1= 3.8129 \ ,\ N_s^2=  3.7502\ ,\ N_s^3= -1.1896 \ , g_s= 0.0049920 \ ,\nn\\
&r^1= 21.330  \ ,\ r^2=  139.02 \ ,\ r^3= 48.433 \ ,\ r^4= 18.181 \ ,\ r^5= 17.040 \ ,\ r^6= 10.420\ ,\nn\\
& N_{1\, 5}= 1.0495 \ , \ N_{H\, 125} = -6.5720 \ ,\ N_{H\, 126} = -2.5142 \ , \ N_{H\, 345} = 1.1946 \ ,\ N_{H\, 346} = -2.2290 \ , \nn\\
& N_{3\, 135} = 535.90 \ ,\ N_{3\, 136} = -197.23 \ ,\ N_{3\, 146} = -20.407 \ ,\ N_{3\, 235} = 46.069 \ ,\nn\\
& N_{3\, 236} = -1056.5 \ ,\ N_{3\, 245} = 581.73 \ ,\ N_{3\, 246} =  591.19 \ .\nn
\eea
\bea
& {\rm Solution}\ 16\ (\mbox{with}\ \lambda= 2240.8)\!:\\
& N_s^1= 15.679 \ ,\ N_s^2= 14.954   \ ,\ N_s^3= - 1.1373  \ , g_s= 0.043724 \ ,\nn\\
&r^1= 226.94  \ ,\ r^2= 30.146 \ ,\ r^3= 33.202 \ ,\ r^4= 23.017 \ ,\ r^5= 260.92\ ,\ r^6= 10.358\ ,\nn\\
& N_{1\, 5}=  -1.0202 \ , \ N_{H\, 125} = 31.252\ ,\ N_{H\, 126} = -2.9713\ , \ N_{H\, 345} = -1.1160\ ,\ N_{H\, 346} = 1.0383 \ , \nn\\
& N_{3\, 136} =  280.63 \ ,\ N_{3\, 235} =  1356.3 \ , \ N_{3\, 236} = 110.62 \ ,\ N_{3\, 246} = 25.843  \ .\nn
\eea
\bea
& {\rm Solution}\ 17\ (\mbox{with}\ \lambda= 621.44)\!:\\
& N_s^1= 15.389 \ ,\ N_s^2= 14.582 \ ,\ N_s^3=  -1.1706 \ , g_s=  0.051272 \ ,\nn\\
&r^1= 121.54\ ,\ r^2= 15.696 \ ,\ r^3= 11.537 \ ,\ r^4= 19.689 \ ,\ r^5= 77.044 \ ,\ r^6= 10.446\ ,\nn\\
& N_{1\, 5}=  1.0352 \ , \ N_{H\, 125} = -25.948 \ ,\ N_{H\, 126} = -3.7303 \ , \ N_{H\, 345} =  -1.1411 \ ,\nn\\
&N_{H\, 346} = -1.2029 \ , \  N_{3\, 136} = -135.73  \ ,\ N_{3\, 235} =  319.44 \ , \ N_{3\, 236} = - 54.882 \ ,\ N_{3\, 246} = 29.915 \ .\nn
\eea
From these values, we observe some internal hierarchies among the radii, as expected \cite{Andriot:2019wrs}. Some hierarchies are also present in the flux ``integers'', in agreement with \cite{Junghans:2018gdb}. The solution 12, that allows $N_s^I <4$, is particularly interesting.

As mentioned previously, it does not make sense to check further the requirements \eqref{requir} for the solution to be in the classical regime, without a detailed knowledge of the 6d geometry. In particular, identifying explicitly the lattice of the group manifold is only achieved for solutions 14 and 15, as explained in section \ref{sec:compactness}. For the other solutions, we cannot check constraints on the $f^a{}_{bc}$ and $N_{abc}$, or determine precisely the upper bound on $N_s^I$. As a change of basis on the forms is probably needed, it is also not worth trying to obtain integers for $\{N_{q\, a_1 \dots a_q}, N_s^I \}$. Finally, quantizing only the harmonic flux components would actually reduce the number of constraints, but those cannot be determined for now. So we stop here the test of whether the de Sitter solutions found are in a classical string regime: we conclude positively for 4 solutions, regarding the requirements \eqref{constraintsNs} with $N_s^I \leq 16$. This study already goes beyond what has been done so far in the literature in 4d frameworks, and provides an interesting illustration of the idea of the ``grey zone'' discussed at the beginning of this section. A complete analysis will be performed on solutions 14 and 15 in the companion paper \cite{Andriot:2020vlg}.

\section{Summary and outlook}\label{sec:ccl}

In this paper, we have found and studied 17 new de Sitter solutions of 10d type IIB supergravity with intersecting $D_5$-branes and $O_5$ orientifold planes, on 6d group manifolds. The solutions were found numerically, following a method described in section \ref{sec:dSsol} and appendix \ref{ap:num}, allowing us to solve the 10d equations and constraints introduced in section \ref{sec:framework}. A last constraint, the compactness of the 6d group manifold, remains difficult to verify as explained in section \ref{sec:compactness} and appendix \ref{ap:alg}, but we could establish it successfully for 4 solutions. The 17 solutions are listed explicitly in appendix \ref{ap:sol}. Such de Sitter solutions were expected \cite{Andriot:2017jhf} because of the formal similarity with the type IIA setting with intersecting $O_6/D_6$, where most de Sitter solutions were previously found \cite{Caviezel:2008tf, Flauger:2008ad, Danielsson:2010bc, Danielsson:2011au, Roupec:2018mbn}. However, the solutions found here remain truly new, in the sense that they are not T-dual to the latter, nor to the only other 10d type IIB supergravity de Sitter solutions known, with $O_5/O_7$ sources \cite{Caviezel:2009tu}. Our solutions could rather be T-dual to other, yet undiscovered, de Sitter solutions having intersecting $O_4/D_4$ and $O_6/D_6$ sources. Finally, our search for solutions accidentally led to the discovery of a new Minkowski solution with intersecting $O_5/D_5$, discussed in section \ref{sec:Mink}.

All our de Sitter solutions were found to be perturbatively unstable. To show this, we analysed a corresponding 4d scalar potential depending only on four fields, $V(\rho,\tau,\sun,\sde)$, building on \cite{Danielsson:2012et, Junghans:2016uvg, Andriot:2018ept, Andriot:2019wrs, Andriot:2020lea}. A tachyonic direction was always found in this 4-field space at the de Sitter critical point, as explained in section \ref{sec:stab}. The details of the potential and the scalar fields kinetic terms are given in section \ref{sec:pot} and appendix \ref{ap:4d}. Thanks to those we computed the values of $\eta_V$ at the critical points, and summarized them in Table \ref{tab:eta}. They are in agreement with the swampland refined de Sitter conjecture of \cite{Ooguri:2018wrx}. Further comments on these values and a related mathematical lemma are given in section \ref{sec:lemma}.

Finally, an important question is whether our 10d supergravity de Sitter solutions are classical string backgrounds. A list of corresponding requirements on the 10d solution is introduced and discussed in section \ref{sec:settingquantiz}, including e.g.~a small string coupling, $g_s \ll 1$, or large internal 6d radii compared to the string length, $r^a \gg l_s$, etc.. Testing some of these conditions however requires a detailed knowledge of the 6d geometry, namely the flux quantization, the count of the number of orientifold planes, and the lattice quantization conditions. For now, we only have a good understanding of the geometry for our solutions 14 and 15, this difficulty is related to the matter of compactness of the group manifold. In section \ref{sec:firstcheckquantiz}, we then test the validity of our de Sitter solutions as classical string backgrounds against a less complete list of requirements, while still going beyond what has been done previously in the literature. In particular, we show that allowing for six different radii, instead of a single volume modulus, provides a useful flexibility to solve the constraints. As a result, 4 of our de Sitter solutions pass successfully these first tests, and are thus good candidates of classical de Sitter string backgrounds. Interestingly, these solutions exhibit internal hierarchies among their radii, as well as their flux integers. A complete study of the solutions 14 and 15 is delayed to the companion paper \cite{Andriot:2020vlg}.\\

Several aspects of this work call for further investigations. To start with, a more thorough search for solutions could be pursued, with possible improvements on our numerical methods. We could also get some inspiration from our solutions to search with a sharper ansatz. In particular, having few structure constants remains an important advantage when dealing with the compactness issue and the identification of the 6d geometry. In addition, it would be interesting to identify the non-solvable algebras appearing in our solutions, and test the compactness of their 6d manifold, as explained in section \ref{sec:compactness}. Solutions 10 and 12 would be prime targets for this, because they appeared as promising candidates of classical de Sitter solutions in section \ref{sec:firstcheckquantiz}. We also note that, as discussed in \cite{Danielsson:2011au, Andriot:2018wzk, Andriot:2018ept}, we have not found any de Sitter solution on a nilmanifold (see however the Minkowski solution of section \ref{sec:Mink}), a point that could be worth understanding better. Other aspects are the lessons learned from our stability study, using the 4-field potential $V(\rho,\tau,\sun,\sde)$. This could be useful to a generic identification of the tachyon, possibly joining previous proposals \cite{Covi:2008ea, Danielsson:2012et, Junghans:2016uvg, Junghans:2016abx}. It may also allow to obtain a general bound on $\eta_V$, that should be compared to several swampland conjectures \cite{Ooguri:2018wrx, Gautason:2018gln, Lust:2019zwm}. More generally, it would interesting to study whether our setting can be subject to instabilities recently discussed in \cite{GarciaEtxebarria:2020xsr}. Finally, the discussion on the classical regime of string theory highlights various subtleties on this topic. The flexibility offered by having different radii instead of a single volume modulus motivates a further study of the known type IIA de Sitter solutions, previously analysed in that respect in \cite{Roupec:2018mbn, Junghans:2018gdb, Banlaki:2018ayh}. In 4d approaches, this would require to include more scalar fields; alternatively, one could aim at reproducing the analysis in the 10d language used here. Our discussion also emphasized the difference between an asymptotic limit in field space or a parametric control, versus a ``grey zone'' where fields remain finite but take large/small enough values to accommodate a classical regime. De Sitter solutions might be forbidden in the former, but some of our solutions may lie in the latter field space region. We note that this could be enough for relevant physics, since some fields, even though large, admit observational upper bounds, as for instance the 6d radii.

Beyond the question of the classical regime, or the verification of compactness, the main criticism against our solutions is a standard one on intersecting sources: they are ``smeared'' \cite{Blaback:2010sj, Junghans:2013xza, Baines:2020dmu}.\footnote{Contrary to $D$-branes, smearing $O$-planes is particularly prohibited, since by definition, an $O$-plane stands at a fixed point. It is worth noting that our solution 14 is special since, as discussed around \eqref{lambda1}, it is the only one with $T_{10}^2 <0$. This means that it can be interpreted as having $O_5$-planes only along the set $I=1$, i.e.~directions (12). It could then be interesting to localize first these sources, while those along orthogonal directions could remain as smeared branes.} We prefer to view our solutions as solving an integrated version of the equations, which trades functions (warp factor, dilaton) and distributions (source $\delta$-functions) for constants; see a discussion in \cite{Andriot:2018wzk}. The question remains whether a localized version exists; it would capture the backreaction of our $O_5/D_5$. This is a well-known supergravity problem \cite{Lu:1997mi, Youm:1999zs, Cvetic:2000cj, Smith:2002wn}: while localized solutions exist for parallel sources, e.g.~\cite{Andriot:2019hay}, they are most of the time unknown for intersecting ones (see however \cite{Assel:2011xz, Assel:2012cj, Rota:2015aoa} in anti-de Sitter). It is also unclear whether finding a localized solution in supergravity is relevant: indeed, the backreaction of sources is a priori important only close to them, where stringy contributions should also be taken into account. In any case, it is often believed that this problem could be cured in full string theory. Interestingly, this question has reappeared recently in the context of the swampland, with the conjectures \cite{Gautason:2018gln, Lust:2019zwm}. The anti-de Sitter solution \cite{DeWolfe:2005uu} is a counter-example to the latter. The main criticism against this solution is again its non-localized intersecting sources. This has motivated a (partial) localization of this anti-de Sitter solution \cite{Junghans:2020acz, Marchesano:2020qvg}. It would be very interesting to study our de Sitter solutions with intersecting sources in this new light.

Finally, it could be interesting to add anti-$D_5$-branes to our setting. Deforming this way our solutions, one can hope to find another tachyonic de Sitter extremum of the potential, and close to it, a de Sitter minimum. This program indeed worked in the example considered in \cite{Kallosh:2018nrk}, that only had few scalar fields. The intuition behind this idea is that staying close enough to the tachyonic point maintains the other directions stabilised, while a tuned $\bar{D}_5$ contribution to the potential can generate a minimum along the tachyonic direction. Given this new de Sitter minimum, a remaining step would then be to verify that it is a solution to the 10d equations, in particular to the flux Bianchi identities. Each step in this program is nevertheless difficult, and we have not succeeded for now, starting with our de Sitter solutions. One complication in our setting is the presence of $D_5$ along directions (56): $\bar{D}_5$ can then not be added there without triggering an instability. A way out could be that the deformed solution admits $T_{10}^3 = 0$, but this remains unlikely given that we have not found any such de Sitter solution. We hope to come back to this idea in future work.

Despite having unstable de Sitter solutions with $\eta_V < -1$, it could still be interesting to construct from them cosmological models. In multi-field inflation, one can actually construct viable models in such situations when allowing for non-geodesic motion or strong bending in field space (see e.g.~\cite{Brown:2017osf, Garcia-Saenz:2018ifx, Achucarro:2018vey, Bjorkmo:2019aev, Bjorkmo:2019fls}). To that end, having a more complete 4d theory from our 10d setting than the one considered in this work could be useful, e.g.~to have a better control on the 4d mass spectrum. A good starting point for that could be the supergravity theory considered for the de Sitter solution with $O_5/O_7$ \cite{Caviezel:2009tu}. We hope to come back to these interesting questions in the future.

\vspace{0.4in}

\subsection*{Acknowledgements}

We warmly thank N.~Cribiori, S.~Renaux-Petel, C.~Roupec, C.~Ruef, H.~Skarke and D.~Tsimpis for useful exchanges during the completion of this work. D.~A.~and P.~M.~acknowledge support from the Austrian Science Fund (FWF): project number M2247-N27. P.~M.~thanks the ITP at TU Wien for hospitality and for the opportunity to work on this project. T.~W.~acknowledges support from the Austrian Science Fund (FWF): project number P 30265.

\newpage

\begin{appendix}

\section{List of de Sitter solutions}\label{ap:sol}

We give in this appendix the explicit list of 17 de Sitter solutions found in this work, and discussed in section \ref{sec:dSsol}. The numerical values displayed correspond to quantities in units of $2\pi l_s$, as explained around \eqref{adim}. To ease the use of these solutions, e.g.~to manipulate them in Mathematica, we hide the dependence on $g_s$ and the indices of the components are put inside square brackets: $T_{10}[I]$ stands for $g_s\, T_{10}^I$, $F_1[a]$ for $g_s\, F_{1\, a}$, $F_3[a,b,c]$ for $g_s\, F_{3\, abc}$, $H[a,b,c]$ for $H_{abc}$ and $f[a,b,c]$ for $\f{a}{bc}$. The values are rounded to 5 significant digits (if less are displayed this means that the following digits are zero in the rounded value), except for the interesting solutions 10, 12, 14, 15, 16, 17, which are displayed with 16 digits. Only the non-vanishing variables are given. For each solution, we also provide the values of $\Rc_4$ and $\Rc_6$, the 4 eigenvalues of the mass matrix $g^{ik}H_{kj}$ for the 4-field potential $V(\rho, \tau, \sun, \sde)$, and the tachyonic eigenvector $\vec{v}$ associated to the negative eigenvalue (see section \ref{sec:stabana}). \\ \\

\vspace{0.4in}

\subsection*{Solution 1}
\label{sol1}

\vspace{0.15in}

\begin{equation*}
\begin{aligned}
&T_{10}[1]\rightarrow 0.47704,T_{10}[2]\rightarrow 0.30751,T_{10}[3]\rightarrow -0.053848,F_1[5]\rightarrow 0.067964,F_1[6]\rightarrow -0.16337,\\[6pt]
&F_3[1,3,5]\rightarrow 0.029423,F_3[1,3,6]\rightarrow 0.042531,F_3[1,4,5]\rightarrow 0.071507,F_3[1,4,6]\rightarrow 0.25908,\\[6pt]
&F_3[2,3,5]\rightarrow -0.0029428,F_3[2,3,6]\rightarrow -0.011609,F_3[2,4,5]\rightarrow -0.026656, F_3[2,4,6]\rightarrow  -0.056824,\\[6pt]
&H[1,2,5]\rightarrow -0.089255,H[1,2,6]\rightarrow -0.020459,H[3,4,5]\rightarrow -0.10652,
H[3,4,6]\rightarrow -0.0097439,\\[6pt]
&f[1,3,5]\rightarrow -0.20034,f[1,3,6]\rightarrow -0.019633,f[1,4,5]\rightarrow 0.075638,
f[1,4,6]\rightarrow -0.060835,\\[6pt]
&f[2,3,5]\rightarrow 0.078361,f[2,3,6]\rightarrow -0.045623,f[2,4,5]\rightarrow -0.019626,
f[2,4,6]\rightarrow 0.012723,\\[6pt]
&f[3,1,5]\rightarrow 0.0062025,f[3,1,6]\rightarrow -0.013864,f[3,2,5]\rightarrow 0.042311,
f[3,2,6]\rightarrow -0.061633,\\[6pt]
&f[4,1,5]\rightarrow 0.016689,f[4,1,6]\rightarrow -0.051754,f[4,2,5]\rightarrow -0.16991,
f[4,2,6]\rightarrow -0.037573,\\[6pt]
&f[5,1,3]\rightarrow -0.0012195,f[5,1,4]\rightarrow 0.00029436,f[5,2,3]\rightarrow -0.046738,
f[5,2,4]\rightarrow 0.011281,\\[6pt]
&f[6,1,3]\rightarrow -0.00050733,f[6,1,4]\rightarrow 0.00012246,f[6,2,3]\rightarrow -0.019444,
f[6,2,4]\rightarrow 0.0046932 \,.
\end{aligned}
\end{equation*}

\vspace{0.15in}

\begin{equation*}
\Rc_4 = 0.011482 \,, \; \Rc_6 = -0.062461 \,,
\end{equation*}

\vspace{-0.1in}

\begin{equation*}
\text{masses}^2 = (0.086586,0.048623,0.044431,-0.0081936) \,, \; \vec{v} = (0.512,0.83441,0.15942,0.12727) \,.
\end{equation*}

\newpage

\subsection*{Solution 2}
\label{sol2}
\begin{equation*}
\begin{aligned}
&T_{10}[1]\rightarrow 0.46469,T_{10}[2]\rightarrow 0.4183,T_{10}[3]\rightarrow -0.13527,F_1[5]\rightarrow -0.054338,F_1[6]\rightarrow 0.09419,\\[6pt]
&F_3[1,3,5]\rightarrow 0.19696,F_3[1,3,6]\rightarrow 0.029077,F_3[1,4,5]\rightarrow -0.14439,F_3[1,4,6]\rightarrow -0.018361,\\[6pt]
&F_3[2,3,5]\rightarrow -0.14504,F_3[2,3,6]\rightarrow -0.023914,F_3[2,4,5]\rightarrow 0.1403,F_3[2,4,6]\rightarrow 0.01418,\\[6pt]
&H[1,2,5]\rightarrow -0.0061166,H[1,2,6]\rightarrow 0.019678,H[3,4,5]\rightarrow 0.0072392,H[3,4,6]\rightarrow 0.029453,\\[6pt]
&f[1,3,5]\rightarrow -0.043274,f[1,3,6]\rightarrow -0.12877,f[1,4,5]\rightarrow -0.0053473,f[1,4,6]\rightarrow -0.19337,\\[6pt]
&f[2,3,5]\rightarrow 0.11672,f[2,3,6]\rightarrow 0.095978,f[2,4,5]\rightarrow 0.12856,f[2,4,6]\rightarrow 0.073781,\\[6pt]
&f[3,1,5]\rightarrow -0.075441,f[3,1,6]\rightarrow 0.10485,f[3,2,5]\rightarrow -0.017193,f[3,2,6]\rightarrow 0.10796, \\[6pt]
&f[4,1,5]\rightarrow 0.070637,f[4,1,6]\rightarrow -0.11098,f[4,2,5]\rightarrow 0.033071,f[4,2,6]\rightarrow -0.091882,\\[6pt]
&f[5,1,3]\rightarrow -0.028108,f[5,1,4]\rightarrow -0.027949,f[5,2,3]\rightarrow 0.084778,f[5,2,4]\rightarrow 0.0843,\\[6pt]
&f[6,1,3]\rightarrow -0.016215,f[6,1,4]\rightarrow -0.016124,f[6,2,3]\rightarrow 0.048908,f[6,2,4]\rightarrow 0.048633 \,.
\end{aligned}
\end{equation*}
\begin{equation*}
\Rc_4 = 0.01048 \,, \;  \Rc_6 = -0.072118 \,,
\end{equation*}
\begin{equation*}
\text{masses}^2 = (0.12192,0.054898,0.02209,-0.0070818) \,, \;  \vec{v} = (0.59801,0.78544,0.083651,0.13587) \,.
\end{equation*}

\subsection*{Solution 3}
\label{sol3}
\begin{equation*}
\begin{aligned}
&T_{10}[1]\rightarrow 0.72554,T_{10}[2]\rightarrow 0.56275,T_{10}[3]\rightarrow -0.11084,F_1[5]\rightarrow -0.046697,F_1[6]\rightarrow -0.14894,\\[6pt]
&F_3[1,3,5]\rightarrow -0.3778,F_3[1,3,6]\rightarrow -0.053076,F_3[1,4,5]\rightarrow 0.078924,F_3[1,4,6]\rightarrow -0.0016181,\\[6pt]
&F_3[2,3,5]\rightarrow 0.0067415,F_3[2,3,6]\rightarrow -0.015912,F_3[2,4,5]\rightarrow -0.0036507,F_3[2,4,6]\rightarrow -0.003343,\\[6pt]
&H[1,2,5]\rightarrow -0.011276,H[1,2,6]\rightarrow 0.046065,H[3,4,5]\rightarrow -0.028408,H[3,4,6]\rightarrow 0.039204,\\[6pt]
&f[1,3,5]\rightarrow 0.036555,f[1,3,6]\rightarrow 0.1166,f[1,4,5]\rightarrow -0.13275,f[1,4,6]\rightarrow 0.25613,\\[6pt]
&f[2,3,5]\rightarrow -0.0070935,f[2,3,6]\rightarrow -0.022625,f[2,4,5]\rightarrow -0.075855,f[2,4,6]\rightarrow -0.077621,\\[6pt]
&f[3,1,5]\rightarrow 0.055343,f[3,1,6]\rightarrow 0.077441,f[3,2,5]\rightarrow 0.08688,f[3,2,6]\rightarrow -0.23348,\\[6pt]
&f[4,1,5]\rightarrow -0.012378,f[4,1,6]\rightarrow -0.0034012,f[4,2,5]\rightarrow -0.06379,f[4,2,6]\rightarrow -0.017528,\\[6pt]
&f[5,1,4]\rightarrow -0.009863,f[5,2,4]\rightarrow -0.050828,f[6,1,4]\rightarrow 0.0030922,f[6,2,4]\rightarrow 0.015935 \,.
\end{aligned}
\end{equation*}
\begin{equation*}
\Rc_4 = 0.019772 \,, \;  \Rc_6 = -0.1156 \,,
\end{equation*}
\begin{equation*}
\text{masses}^2 = (0.1202,0.074661,0.042293,-0.014499) \,, \;  \vec{v} = (0.5718,0.79714,0.1371,0.13717) \,.
\end{equation*}

\subsection*{Solution 4}
\label{sol4}
\begin{equation*}
\begin{aligned}
&T_{10}[1]\rightarrow 0.35972,T_{10}[2]\rightarrow 0.52854,T_{10}[3]\rightarrow -0.061607,F_1[5]\rightarrow -0.055616,F_1[6]\rightarrow -0.089616,\\[6pt]
&F_3[1,3,5]\rightarrow -0.2955,F_3[1,3,6]\rightarrow 0.14475,F_3[1,4,5]\rightarrow 0.053224,F_3[1,4,6]\rightarrow -0.01737,\\[6pt]
&F_3[2,3,5]\rightarrow 0.036808,F_3[2,3,6]\rightarrow -0.033252,F_3[2,4,5]\rightarrow -0.043811,F_3[2,4,6]\rightarrow 0.014833,\\[6pt]
&H[1,2,5]\rightarrow -0.0044838,H[1,2,6]\rightarrow 0.014822,H[3,4,5]\rightarrow 0.01242,H[3,4,6]\rightarrow 0.0093064,\\[6pt]
&f[1,3,5]\rightarrow 0.021656,f[1,3,6]\rightarrow 0.044792,f[1,4,5]\rightarrow -0.026067,f[1,4,6]\rightarrow 0.2064,\\[6pt]
&f[2,3,5]\rightarrow -0.018528,f[2,3,6]\rightarrow -0.033604,f[2,4,5]\rightarrow 0.05014,f[2,4,6]\rightarrow -0.013342,\\[6pt]
&f[3,1,5]\rightarrow -0.057579,f[3,1,6]\rightarrow 0.011876,f[3,2,5]\rightarrow -0.18459,f[3,2,6]\rightarrow -0.18172,\\[6pt]
&f[4,1,5]\rightarrow 0.0086028,f[4,1,6]\rightarrow 0.040689,f[4,2,5]\rightarrow 0.01433,f[4,2,6]\rightarrow 0.052753,\\[6pt]
&f[5,1,3]\rightarrow -0.00083758,f[5,1,4]\rightarrow -0.021022,f[5,2,3]\rightarrow -0.0009261,f[5,2,4]\rightarrow -0.023244,\\[6pt]
&f[6,1,3]\rightarrow 0.00051981,f[6,1,4]\rightarrow 0.013047,f[6,2,3]\rightarrow 0.00057474,f[6,2,4]\rightarrow 0.014425 \,.
\end{aligned}
\end{equation*}
\begin{equation*}
\Rc_4 = 0.010644 \,, \;  \Rc_6 = -0.079291 \,,
\end{equation*}
\begin{equation*}
\text{masses}^2 = (0.081167,0.046349,0.018831,-0.0077075) \,, \;  \vec{v} = (0.59942,0.77513,0.12971,0.15182) \,.
\end{equation*}

\subsection*{Solution 5}
\label{sol5}
\begin{equation*}
\begin{aligned}
&T_{10}[1]\rightarrow 0.5488,T_{10}[2]\rightarrow 0.49801,T_{10}[3]\rightarrow -0.09235,F_1[5]\rightarrow -0.1363,F_1[6]\rightarrow -0.035535,\\[6pt]
&F_3[1,3,5]\rightarrow -0.058902,F_3[1,3,6]\rightarrow -0.32976,F_3[1,4,5]\rightarrow 0.011126,F_3[1,4,6]\rightarrow 0.1021,\\[6pt]
&F_3[2,3,5]\rightarrow -0.0067774,F_3[2,3,6]\rightarrow 0.0038507,F_3[2,4,6]\rightarrow -0.005906,F_3[2, 4, 5] \rightarrow  F_3[2, 4, 6],\\[6pt]
&H[1,2,5]\rightarrow -0.041183,H[1,2,6]\rightarrow 0.015758,H[3,4,5]\rightarrow -0.034816,H[3,4,6]\rightarrow 0.018389,\\[6pt]
&f[1,3,5]\rightarrow -0.12993,f[1,3,6]\rightarrow -0.033873,f[1,4,5]\rightarrow -0.20816,f[1,4,6]\rightarrow 0.095049,\\[6pt]
&f[2,3,5]\rightarrow 0.025631,f[2,3,6]\rightarrow 0.0066821,f[2,4,6]\rightarrow 0.04532,f[3,1,5]\rightarrow -0.054525,\\[6pt]
&f[3,2,5]\rightarrow 0.22297,f[3,2,6]\rightarrow -0.099548,f[4,1,5]\rightarrow 0.0012519,f[4,1,6]\rightarrow 0.018843,\\[6pt]
&f[4,2,5]\rightarrow 0.006346,f[4,2,6]\rightarrow 0.095521,f[5,1,4]\rightarrow 0.0025089,f[5,2,4]\rightarrow 0.012718,\\[6pt]
&f[6,1,4]\rightarrow -0.0096234,f[6,2,4]\rightarrow -0.048783,f[2, 4, 5] \rightarrow  f[2, 4, 6], f[3, 1, 6] \rightarrow  -f[2, 4, 6] \,.
\end{aligned}
\end{equation*}
\begin{equation*}
\Rc_4 = 0.016346 \,, \;  \Rc_6 = -0.094138 \,,
\end{equation*}
\begin{equation*}
\text{masses}^2 = (0.091114,0.065509,0.03436,-0.012138) \,, \;  \vec{v} = (0.56953,0.79882,0.14606,0.12724) \,.
\end{equation*}

\subsection*{Solution 6}
\label{sol6}
\begin{equation*}
\begin{aligned}
&T_{10}[1]\rightarrow 0.18633,T_{10}[2]\rightarrow 0.099822,T_{10}[3]\rightarrow -0.023249,F_1[5]\rightarrow -0.06428,F_1[6]\rightarrow -0.034831,\\[6pt]
&F_3[1,3,5]\rightarrow 0.01078,F_3[1,3,6]\rightarrow -0.026777,F_3[1,4,5]\rightarrow -0.016182,F_3[1,4,6]\rightarrow 0.033328,\\[6pt]
&F_3[2,3,5]\rightarrow -0.015359,F_3[2,3,6]\rightarrow 0.10935,F_3[2,4,5]\rightarrow -0.0041419,F_3[2,4,6]\rightarrow -0.14142,\\[6pt]
&H[1,2,5]\rightarrow -0.024728,H[1,2,6]\rightarrow -0.0016963,H[3,4,5]\rightarrow -0.020168,H[3,4,6]\rightarrow 0.016339,\\[6pt]
&f[1,3,5]\rightarrow -0.043546,f[1,3,6]\rightarrow -0.03002,f[1,4,5]\rightarrow -0.040593,f[1,4,6]\rightarrow -0.024861,\\[6pt]
&f[2,3,5]\rightarrow 0.10185,f[2,3,6]\rightarrow -0.050296,f[2,4,5]\rightarrow 0.10301,f[2,4,6]\rightarrow 0.0087766,\\[6pt]
&f[3,1,5]\rightarrow 0.03814,f[3,1,6]\rightarrow -0.032997,f[3,2,5]\rightarrow -0.005429,f[3,2,6]\rightarrow -0.022669,\\[6pt]
&f[4,1,5]\rightarrow -0.09408,f[4,1,6]\rightarrow 0.0059857,f[4,2,5]\rightarrow 0.0090268,f[4,2,6]\rightarrow 0.025832,\\[6pt]
&f[5,1,3]\rightarrow 0.0098816,f[5,1,4]\rightarrow 0.0044069,f[5,2,3]\rightarrow 0.0036326,f[5,2,4]\rightarrow 0.0016201,\\[6pt]
&f[6,1,3]\rightarrow -0.018236,f[6,1,4]\rightarrow -0.008133,f[6,2,3]\rightarrow -0.006704,f[6,2,4]\rightarrow -0.0029898 \,.
\end{aligned}
\end{equation*}
\begin{equation*}
\Rc_4 =0.004057 \,, \;  \Rc_6 = -0.025322 \,,
\end{equation*}
\begin{equation*}
\text{masses}^2 = (0.030262,0.013323,0.0092559,-0.0029562) \,, \;  \vec{v} = (0.56648,0.80047,0.14645,0.12997) \,.
\end{equation*}

\subsection*{Solution 7}
\label{sol7}
\begin{equation*}
\begin{aligned}
&T_{10}[1]\rightarrow 0.32241,T_{10}[2]\rightarrow 0.25405,T_{10}[3]\rightarrow -0.1001,F_1[5]\rightarrow 0.027908,F_1[6]\rightarrow 0.079651,\\[6pt]
&F_3[1,3,5]\rightarrow 0.057728,F_3[1,3,6]\rightarrow -0.0003245,F_3[1,4,5]\rightarrow -0.11383,F_3[1,4,6]\rightarrow 0.0043009,\\[6pt]
&F_3[2,3,5]\rightarrow 0.12422,F_3[2,3,6]\rightarrow 0.015909,F_3[2,4,5]\rightarrow -0.18375,F_3[2,4,6]\rightarrow -0.010172,\\[6pt]
&H[1,2,5]\rightarrow -0.0042465,H[1,2,6]\rightarrow -0.014838,H[3,4,5]\rightarrow 0.014738,H[3,4,6]\rightarrow -0.016011,\\[6pt]
&f[1,3,5]\rightarrow -0.0078082,f[1,3,6]\rightarrow -0.14573,f[1,4,5]\rightarrow 0.01363,f[1,4,6]\rightarrow -0.059324,\\[6pt]
&f[2,3,5]\rightarrow 0.14929,f[2,3,6]\rightarrow -0.10111,f[2,4,5]\rightarrow 0.11852,f[2,4,6]\rightarrow -0.081211,\\[6pt]
&f[3,1,5]\rightarrow -0.036299,f[3,1,6]\rightarrow -0.10329,f[3,2,5]\rightarrow 0.0019146,f[3,2,6]\rightarrow 0.028735,\\[6pt]
&f[4,1,5]\rightarrow 0.045777,f[4,1,6]\rightarrow 0.12971,f[4,2,5]\rightarrow 0.0092216,f[4,2,6]\rightarrow -0.044147,\\[6pt]
&f[5,1,3]\rightarrow 0.0014146,f[5,1,4]\rightarrow 0.0011256,f[5,2,3]\rightarrow 0.1058,f[5,2,4]\rightarrow 0.084183,\\[6pt]
&f[6,1,3]\rightarrow -0.00049565,f[6,1,4]\rightarrow -0.00039437,f[6,2,3]\rightarrow -0.037071,f[6,2,4]\rightarrow -0.029496 \,.
\end{aligned}
\end{equation*}
\begin{equation*}
\Rc_4 =0.0064113 \,, \;  \Rc_6 = -0.045751 \,,
\end{equation*}
\begin{equation*}
\text{masses}^2 = (0.09705,0.033829,0.014147,-0.0040233) \,, \;  \vec{v} = (0.61748,0.7735,0.071634,0.12359) \,.
\end{equation*}

\subsection*{Solution 8}
\label{sol8}
\begin{equation*}
\begin{aligned}
&T_{10}[1]\rightarrow 0.19717,T_{10}[2]\rightarrow 0.14783,T_{10}[3]\rightarrow -0.021473,F_1[5]\rightarrow -0.078205,F_1[6]\rightarrow 0.0049412,\\[6pt]
&F_3[1,3,5]\rightarrow -0.0099514,F_3[1,3,6]\rightarrow 0.011265,F_3[1,4,5]\rightarrow -0.053599,F_3[1,4,6]\rightarrow 0.17075,\\[6pt]
&F_3[2,3,5]\rightarrow 0.018747,F_3[2,3,6]\rightarrow -0.029133,F_3[2,4,5]\rightarrow -0.051305,F_3[2,4,6]\rightarrow 0.085993,\\[6pt]
&H[1,2,5]\rightarrow 0.015719,H[1,2,6]\rightarrow 0.0048415,H[3,4,5]\rightarrow 0.03184,H[3,4,6]\rightarrow 0.023596,\\[6pt]
&f[1,3,5]\rightarrow -0.12703,f[1,3,6]\rightarrow 0.028513,f[1,4,5]\rightarrow 0.012977,f[1,4,6]\rightarrow 0.014053,\\[6pt]
&f[2,3,5]\rightarrow -0.086914,f[2,3,6]\rightarrow -0.033169,f[2,4,5]\rightarrow -0.025341,f[2,4,6]\rightarrow -0.026465,\\[6pt]
&f[3,1,5]\rightarrow -0.011566,f[3,1,6]\rightarrow 0.00064098,f[3,2,5]\rightarrow -0.0058617,f[3,2,6]\rightarrow 0.00062058,\\[6pt]
&f[4,1,5]\rightarrow 0.046748,f[4,1,6]\rightarrow 0.031485,f[4,2,5]\rightarrow -0.077789,f[4,2,6]\rightarrow -0.091042,\\[6pt]
&f[5,1,3]\rightarrow 0.00041274,f[5,1,4]\rightarrow 0.00029964,f[5,2,3]\rightarrow -0.00115,f[5,2,4]\rightarrow -0.00083489,\\[6pt]
&f[6,1,3]\rightarrow 0.0065325,f[6,1,4]\rightarrow 0.0047425,f[6,2,3]\rightarrow -0.018202,f[6,2,4]\rightarrow -0.013214 \,.
\end{aligned}
\end{equation*}
\begin{equation*}
\Rc_4 =0.0042994 \,, \;  \Rc_6 = -0.030339 \,,
\end{equation*}
\begin{equation*}
\text{masses}^2 = (0.03168,0.015949,0.010753,-0.002987) \,, \;  \vec{v} = (0.57994,0.78615,0.14473,0.15715) \,.
\end{equation*}

\subsection*{Solution 9}
\label{sol9}
\begin{equation*}
\begin{aligned}
&T_{10}[1]\rightarrow 0.35204,T_{10}[2]\rightarrow 0.32339,T_{10}[3]\rightarrow -0.12879,F_1[5]\rightarrow 0.05975,F_1[6]\rightarrow -0.058962,\\[6pt]
&F_3[1,3,5]\rightarrow 0.11186,F_3[1,3,6]\rightarrow 0.059735,F_3[1,4,5]\rightarrow -0.21025,F_3[1,4,6]\rightarrow -0.094019,\\[6pt]
&F_3[2,3,5]\rightarrow 0.042689,F_3[2,3,6]\rightarrow 0.030822,F_3[2,4,5]\rightarrow -0.060858,F_3[2,4,6]\rightarrow -0.043124,\\[6pt]
&H[1,2,5]\rightarrow 0.0086458,H[1,2,6]\rightarrow -0.0086346,H[3,4,5]\rightarrow -0.0085054,H[3,4,6]\rightarrow -0.015009,\\[6pt]
&f[1,3,5]\rightarrow -0.1416,f[1,3,6]\rightarrow -0.23025,f[1,4,5]\rightarrow -0.06172,f[1,4,6]\rightarrow -0.10511,\\[6pt]
&f[2,3,5]\rightarrow 0.065963,f[2,3,6]\rightarrow -0.095767,f[2,4,5]\rightarrow 0.06644,f[2,4,6]\rightarrow -0.079327,\\[6pt]
&f[3,1,5]\rightarrow 0.00092451,f[3,1,6]\rightarrow -0.011556,f[3,2,5]\rightarrow -0.061413,f[3,2,6]\rightarrow 0.061127,\\[6pt]
&f[4,1,5]\rightarrow -0.0068416,f[4,1,6]\rightarrow 0.018309,f[4,2,5]\rightarrow 0.1371,f[4,2,6]\rightarrow -0.13586,\\[6pt]
&f[5,1,3]\rightarrow -0.13707,f[5,1,4]\rightarrow -0.061505,f[5,2,3]\rightarrow 0.0067551,f[5,2,4]\rightarrow 0.0030311,\\[6pt]
&f[6,1,3]\rightarrow -0.1389,f[6,1,4]\rightarrow -0.062327,f[6,2,3]\rightarrow 0.0068453,f[6,2,4]\rightarrow 0.0030716 \,.
\end{aligned}
\end{equation*}
\begin{equation*}
\Rc_4 =0.0065997 \,, \;  \Rc_6 = -0.051929 \,,
\end{equation*}
\begin{equation*}
\text{masses}^2 = (0.15427,0.045378,0.015301,-0.0037113) \,, \;  \vec{v} = (0.6478,0.75245,0.047002,0.10938) \,.
\end{equation*}

\subsection*{Solution 10}
\label{sol10}
\begin{equation*}
\begin{aligned}
&T_{10}[1]\rightarrow 10,T_{10}[2]\rightarrow 10,T_{10}[3]\rightarrow -0.3259120382713294,F_1[5]\rightarrow 1,F_1[6]\rightarrow 1,\\[6pt]
&F_3[1,3,5]\rightarrow -0.1349772714306872,F_3[1,3,6]\rightarrow -0.7456107008676475,\\[6pt]
&F_3[1,4,5]\rightarrow 0.4757995474652397,F_3[1,4,6]\rightarrow -0.2227565010250167,\\[6pt]
&F_3[2,3,5]\rightarrow -0.4614346028465642,F_3[2,3,6]\rightarrow 0.0426792857654753,\\[6pt]
&F_3[2,4,5]\rightarrow 0.2860470205462396, F_3[2,4,6]\rightarrow -F_3[2,4,5], H[1,2,5]\rightarrow 0.9376250941912930,
\\[6pt]
&H[1,2,6]\rightarrow -0.2994901148924934,H[3,4,5]\rightarrow 0.6659796477910178,\\[6pt]
&H[3,4,6]\rightarrow -0.7329190688589823,f[1,3,6]\rightarrow -0.1969129428812132,\\[6pt]
&f[1,4,5]\rightarrow -0.2753236733652662,f[1,4,6]\rightarrow 0.3984487144549955,\\[6pt]
&f[2,4,5]\rightarrow -0.0799090201373039,f[2,4,6]\rightarrow 0.3243522244088599,\\[6pt]
&f[3,1,6]\rightarrow -0.2002924300421296,f[3,2,5]\rightarrow 0.4230701843222834,\\[6pt]
&f[3,2,6]\rightarrow -0.0284354195594737,f[4,2,5]\rightarrow -0.1236377141499939,\\[6pt]
&f[4,2,6]\rightarrow 0.5018478206399179,f[6,2,4]\rightarrow -0.0889546337076675, \\[6pt]
&f[1,3,5] \rightarrow f[1,3,6], f[3,1,5] \rightarrow f[3,1,6], f[5,2,4] \rightarrow -f[6,2,4] \,.
\end{aligned}
\end{equation*}
\begin{equation*}
\Rc_4 = 0.05046560105547959 \,, \;  \Rc_6 = -0.7152057317272771 \,,
\end{equation*}
\begin{equation*}
\text{masses}^2 = (5.0863,1.233,0.57277,-0.026378) \,, \;  \vec{v} = (0.50049,0.81331,0.21722,0.20212) \,.
\end{equation*}

\subsection*{Solution 11}
\label{sol11}
\begin{equation*}
\begin{aligned}
&T_{10}[1]\rightarrow 0.72539,T_{10}[2]\rightarrow 0.57602,T_{10}[3]\rightarrow -0.11238,F_1[5]\rightarrow -0.046382,F_1[6]\rightarrow -0.14989,\\[6pt]
&F_3[1,3,5]\rightarrow -0.37962,F_3[1,3,6]\rightarrow -0.054102,F_3[1,4,5]\rightarrow 0.078715,F_3[1,4,6]\rightarrow -0.00035418,\\[6pt]
&F_3[2,3,5]\rightarrow 0.0058597,F_3[2,3,6]\rightarrow -0.014748,F_3[2,4,6]\rightarrow -0.0036669,F_3[2,4,5]\rightarrow F_3[2,4,6],\\[6pt]
&H[1,2,5]\rightarrow -0.011716,H[1,2,6]\rightarrow 0.045795,H[3,4,5]\rightarrow -0.02785,H[3,4,6]\rightarrow 0.039606,\\[6pt]
&f[1,3,5]\rightarrow 0.036064,f[1,3,6]\rightarrow 0.11655,f[1,4,5]\rightarrow -0.13436,f[1,4,6]\rightarrow 0.25498,\\[6pt]
&f[2,3,5]\rightarrow -0.0072143,f[2,3,6]\rightarrow -0.023314,f[2,4,5]\rightarrow -0.075938,f[2,4,6]\rightarrow -0.077789,\\[6pt]
&f[3,1,5]\rightarrow 0.056116,f[3,2,5]\rightarrow 0.086816,f[3,2,6]\rightarrow -0.23712,f[4,1,5]\rightarrow -0.012879,\\[6pt]
&f[4,1,6]\rightarrow -0.0033548,f[4,2,5]\rightarrow -0.064383,f[4,2,6]\rightarrow -0.016771,f[5,1,4]\rightarrow -0.010266,\\[6pt]
&f[5,2,4]\rightarrow -0.05132,f[6,1,4]\rightarrow 0.0031768,f[6,2,4]\rightarrow 0.015881,f[3,1,6]\rightarrow -f[2,4,6] \,.
\end{aligned}
\end{equation*}
\begin{equation*}
\Rc_4 =0.02004 \,, \;  \Rc_6 = -0.11684 \,,
\end{equation*}
\begin{equation*}
\text{masses}^2 = (0.12054,0.076154,0.042724,-0.014706) \,, \;  \vec{v} = (0.5719,0.7971,0.13641,0.13769) \,.
\end{equation*}

\subsection*{Solution 12}
\label{sol12}
\begin{equation*}	
\begin{aligned}
&T_{10}[1]\rightarrow 7.7271771858613970,T_{10}[2]\rightarrow 2.2569215718539540,T_{10}[3]\rightarrow -0.1443613520399306,\\[6pt]
&F_1[5] \rightarrow 1, F_3[1,3,5]\rightarrow 0.4942839801759979,F_3[1,3,6]\rightarrow -0.2974725565593828,\\[6pt]
&F_3[1,4,6]\rightarrow -0.0819919421531531,F_3[2,3,5]\rightarrow 0.0065197489236415,\\[6pt]
&F_3[2,3,6]\rightarrow -0.2445092180276900,F_3[2,4,5]\rightarrow 0.2193161877900493,\\[6pt]
&F_3[2,4,6]\rightarrow 0.3644682080232862,H[1,2,5]\rightarrow -0.4230470081250839,\\[6pt]
&H[1,2,6]\rightarrow -0.2646546977657695,H[3,4,5]\rightarrow 0.2589549673229080,\\[6pt]
&H[3,4,6]\rightarrow -0.7901065711295830,f[1,4,5]\rightarrow -0.1277833635641377,\\[6pt]
&f[1,4,6]\rightarrow -0.5365645111668850,f[2,3,5]\rightarrow -0.4614839479014394,\\[6pt]
&f[2,4,5]\rightarrow -0.1062270206129369,f[2,4,6]\rightarrow -0.0551085436402409,\\[6pt]
&f[3,1,5]\rightarrow -0.0119865546470415,f[3,2,5]\rightarrow 0.2444904911694590,\\[6pt]
&f[4,1,5]\rightarrow -0.0650526285279246,f[4,1,6]\rightarrow -0.2731570906621740,\\[6pt]
&f[6,1,4]\rightarrow 0.0984021196994866, f[3,1,6]\rightarrow -f[2,4,6]\,.
\end{aligned}
\end{equation*}
\begin{equation*}
\Rc_4 = 0.059663051023583824 \,, \;  \Rc_6 = -0.4094726936749863 \,,
\end{equation*}
\begin{equation*}
\text{masses}^2 = (2.7155,0.63285,0.25637,-0.04109) \,, \;  \vec{v} = (0.46387,0.85232,0.19198,0.1467) \,.
\end{equation*}

\subsection*{Solution 13}
\label{sol13}
\begin{equation*}
\begin{aligned}
&T_{10}[1]\rightarrow 10,T_{10}[2]\rightarrow 0.77399,T_{10}[3]\rightarrow -0.20221,F_1[5]\rightarrow 0.82894,F_3[1,3,5]\rightarrow 0.1905,\\[6pt]
&F_3[1,3,6]\rightarrow 0.00375,F_3[1,4,6]\rightarrow -0.0039311,F_3[2,3,5]\rightarrow -0.46675,F_3[2,3,6]\rightarrow -0.57687,\\[6pt]
&F_3[2,4,5]\rightarrow 0.0022733,F_3[2,4,6]\rightarrow 0.67087,H[1,2,5]\rightarrow 0.17259,H[1,2,6]\rightarrow -0.00071525,\\[6pt]
&H[3,4,5]\rightarrow 0.0026425,H[3,4,6]\rightarrow -0.78712,f[1,4,6]\rightarrow -0.19142,f[2,3,5]\rightarrow -0.66446,\\[6pt]
&f[2,4,5]\rightarrow -0.54418,f[2,4,6]\rightarrow 0.46713,f[3,1,5]\rightarrow -0.10243,f[3,1,6]\rightarrow 0.1473,\\[6pt]
&f[3,2,5]\rightarrow -0.041972,f[4,1,6]\rightarrow -0.17985,f[6,1,4]\rightarrow 0.058422 \,.
\end{aligned}
\end{equation*}
\begin{equation*}
\Rc_4 =0.037791 \,, \;  \Rc_6 = -0.5941 \,,
\end{equation*}
\begin{equation*}
\text{masses}^2 = (2.3792,0.59264,0.10413,-0.027888) \,, \;  \vec{v} = (0.41291,0.87252,0.22255,0.13672) \,.
\end{equation*}

\subsection*{Solution 14}
\label{sol14}
\begin{equation*}	
\begin{aligned}
&T_{10}[1]\rightarrow 10,T_{10}[2]\rightarrow -0.0885069318066244,T_{10}[3]\rightarrow -0.7765198126057072,\\[6pt]
&F_1[5]\rightarrow -0.2739820106484752,F_3[1,3,5]\rightarrow -0.5612239678297053,\\[6pt]
&F_3[1,3,6]\rightarrow 0.7199875113561189,F_3[1,4,6]\rightarrow 0.0527969424771896,\\[6pt]
&F_3[2,3,5]\rightarrow 0.6773312203822072,F_3[2,3,6]\rightarrow -0.3132864455247597,\\[6pt]
&F_3[2,4,6]\rightarrow 0.1780541307257305,H[1,2,5]\rightarrow -0.0045785440625781,\\[6pt]
&H[3,4,6]\rightarrow 0.2288818622936161,f[1,4,5]\rightarrow 0.8435712996340920,\\[6pt]
&f[1,4,6]\rightarrow 0.6715419200224235,f[2,3,5]\rightarrow -0.2892985071257778,\\[6pt]
&f[2,4,5]\rightarrow -0.0614203186917094,f[2,4,6]\rightarrow -0.8104719938914240,\\[6pt]
&f[3,1,5]\rightarrow 0.0162126509210115,f[3,2,5]\rightarrow 0.0134334990107109,\\[6pt]
&f[6,1,4]\rightarrow 0.4131042712391767\,.
\end{aligned}
\end{equation*}
\begin{equation*}
\Rc_4 = 0.022658272206244612 \,, \;  \Rc_6 = -0.7577021085288538 \,,
\end{equation*}
\begin{equation*}
\text{masses}^2 = (1.9928,0.24874,0.02597,-0.0096679) \,, \;  \vec{v} = (0.11993,0.95779,0.06937,0.25189) \,.
\end{equation*}

\subsection*{Solution 14${}'$ (see appendix \ref{ap:alg})}
\label{sol14'}
\begin{equation*}	
\begin{aligned}
&T_{10}[1]\rightarrow 10, T_{10} [2]\rightarrow -0.0885069318066244,T_{10}[3]\rightarrow -0.7765198126057072,\\[6pt]
&F_1[5]\rightarrow -0.2739820106484752,H[1,2,5]\rightarrow -0.0045785440625781,\\[6pt]
&H[3,4,5]\rightarrow -0.2875146945871226,H[3,4,6]\rightarrow 0.2288818622936161,\\[6pt]
&F_3[1,3,5]\rightarrow -2.7580681436956810,F_3[1,3,6]\rightarrow 1.0980873478564690,\\[6pt]
&F_3[1,4,5]\rightarrow -8.9169397596149600,F_3[1,4,6]\rightarrow 3.4691323044710350,\\[6pt]
&F_3[2,3,5]\rightarrow 1.0708724813659480,F_3[2,3,6]\rightarrow -0.3132864455247597,\\[6pt]
&F_3[2,4,5]\rightarrow 3.3175634705051610,F_3[2,4,6]\rightarrow -0.8579415960403540,\\[6pt]
&f[1,4,6]\rightarrow 0.6715419200224235,f[2,3,5]\rightarrow -0.2892985071257778,\\[6pt]
&f[3,2,5]\rightarrow 0.0134334990107109,f[6,1,4]\rightarrow 0.4131042712391767,\\[6pt]
&g^{12} \rightarrow 1.2068822060495703, g^{34} \rightarrow -3.3068641863224433, g^{56} \rightarrow 1.2561707236473398
\,.
\end{aligned}
\end{equation*}
\begin{equation*}
\Rc_4 \,, \Rc_6\  {\rm unchanged}\,.
\end{equation*}

\subsection*{Solution 15}
\label{sol15}
\begin{equation*}	
\begin{aligned}
&T_{10}[1]\rightarrow 10,T_{10}[2]\rightarrow 0.4966295777593360,T_{10}[3]\rightarrow -0.1058505594207743,\\[6pt]
&F_1[5]\rightarrow 0.1394435290122775,F_3[1,3,6]\rightarrow 0.0034027792189916,\\[6pt]
&F_3[1,4,6]\rightarrow -0.0003267933126085,F_3[2,3,5]\rightarrow 0.0090828099128681,\\[6pt]
&F_3[2,3,6]\rightarrow -1.1496878282089060,F_3[2,4,5]\rightarrow -0.0048365319588065,\\[6pt]
&F_3[2,4,6]\rightarrow 0.6091051220227705,H[1,2,6]\rightarrow -0.0001845709594129,\\[6pt]
&H[3,4,5]\rightarrow -0.0010888652794866,f[2,3,5]\rightarrow -0.6020820458095239,\\[6pt]
&f[2,4,5]\rightarrow -1.1306855450590460,f[3,1,5]\rightarrow -0.0698547712340311,\\[6pt]
&f[3,1,6]\rightarrow -0.1916598129974718,f[3,2,5]\rightarrow -0.0588533218001099,\\[6pt]
&f[4,1,6]\rightarrow 0.1020574931847763,f[6,1,4]\rightarrow 0.0153448261959478\,.
\end{aligned}
\end{equation*}
\begin{equation*}
\Rc_4 = 0.019443278089360194 \,, \;  \Rc_6 = -0.8853409197705556 \,,
\end{equation*}
\begin{equation*}
\text{masses}^2 = (1.6415,0.057392,0.031507,-0.01426) \,, \;  \vec{v} = (0.60222,0.77078,0.17269,0.11581) \,.
\end{equation*}

\subsection*{Solution 15${}'$ (see appendix \ref{ap:alg})}
\label{sol15'}
\begin{equation*}	
\begin{aligned}
&T_{10}[1]\rightarrow 10,T_{10}[2]\rightarrow 0.4966295777593360,T_{10}[3]\rightarrow -0.1058505594207743,\\[6pt]
&F_1[5]\rightarrow 0.1394435290122775,F_3[1,3,5]\rightarrow -0.0107806592596513,\\[6pt]
&F_3[1,3,6]\rightarrow 1.3680017138866560,F_3[1,4,5]\rightarrow 0.0259862634523308,\\[6pt]
&F_3[1,4,6]\rightarrow -3.2923433827652360,F_3[2,3,5]\rightarrow 0.0090828099128681,\\[6pt]
&F_3[2,3,6]\rightarrow -1.1496878282089060,F_3[2,4,5]\rightarrow -0.0218936788185686,\\[6pt]
&F_3[2,4,6]\rightarrow 2.7681720095568790,
H[1,2,6]\rightarrow -0.0001845709594129,\\[6pt]
&H[3,4,5]\rightarrow -0.0010888652794866,f[2,3,5]\rightarrow -0.6020820458095239,\\[6pt]
&f[3,2,5]\rightarrow -0.0588533218001099,f[4,1,6]\rightarrow 0.1020574931847763,\\[6pt]
&f[6,1,4]\rightarrow 0.0153448261959478,\\[6pt]
&g^{12} \rightarrow 1.1869299658443528, g^{34} \rightarrow 1.8779592464658086
\,.
\end{aligned}
\end{equation*}
\begin{equation*}
\Rc_4 \,, \Rc_6\  {\rm unchanged}\,.
\end{equation*}

\subsection*{Solution 16}
\label{sol16}
\begin{equation*}	
\begin{aligned}
&T_{10}[1]\rightarrow 10,T_{10}[2]\rightarrow 1.0653924581926100,T_{10}[3]\rightarrow -0.2865487441401781,\\[6pt]
&F_1[5]\rightarrow -0.3830798073971085,F_3[2,3,5]\rightarrow 0.5088261106821323,\\[6pt]
&F_3[2,3,6]\rightarrow 1.0453549102326770,F_3[2,4,6]\rightarrow 0.3522836755847258,F_3[1,3,6]\rightarrow F_3[2,4,6],\\[6pt]
&H[1,2,5]\rightarrow 0.0392319041342279,H[1,2,6]\rightarrow -0.0939555787571918,\\[6pt]
&H[3,4,5]\rightarrow -0.0125416177756354,H[3,4,6]\rightarrow 0.2939059636978374,\\[6pt]
&f[2,3,5]\rightarrow -0.3584729155627473,f[2,4,5]\rightarrow 0.9572822432446080,\\[6pt]
&f[2,4,6]\rightarrow -0.5911827525101534,f[3,1,5]\rightarrow 0.2190447474382595,\\[6pt]
&f[3,1,6]\rightarrow 0.1889887396788805,f[4,1,5]\rightarrow 0.1145962804793664,\\[6pt]
&f[6,1,4]\rightarrow -0.0456860378765839, f[3,2,5]\rightarrow -f[4,1,5], f[1,4,5]\rightarrow -f[2,3,5]
\,.
\end{aligned}
\end{equation*}
\begin{equation*}
\Rc_4 = 0.0498453380802234 \,, \;  \Rc_6 = -0.8996299138736688 \,,
\end{equation*}
\begin{equation*}
\text{masses}^2 = (1.6235,0.26174,0.12567,-0.035395) \,, \;  \vec{v} = (0.48957,0.83657,0.20509,0.13567) \,.
\end{equation*}

\subsection*{Solution 17}
\label{sol17}
\begin{equation*}	
\begin{aligned}
&T_{10}[1]\rightarrow 10,T_{10}[2]\rightarrow 1.1283773832265060,T_{10}[3]\rightarrow -0.3209376228220143,\\[6pt]
&F_1[5]\rightarrow 0.4281135811392881,F_3[2,3,5]\rightarrow 0.7295336312279515,\\[6pt]
&F_3[2,3,6]\rightarrow -0.9244213776567620,F_3[2,4,6]\rightarrow 0.2952514287937376,
F_3[1,3,6]\rightarrow -F_3[2,4,6],\\[6pt]
&H[1,2,5]\rightarrow -0.1097182799067921,H[1,2,6]\rightarrow -0.1163339858329525,\\[6pt]
&H[3,4,5]\rightarrow -0.0405195014515629,H[3,4,6]\rightarrow -0.3150079434109206,\\[6pt]
&f[2,3,5]\rightarrow -0.3272913073142077,f[2,4,5]\rightarrow -0.8339260055526270,\\[6pt]
&f[2,4,6]\rightarrow -0.7780888837446737,f[3,1,6]\rightarrow -0.2873520591201465,\\[6pt]
&f[4,1,5]\rightarrow -0.1208702926537678,f[6,1,4]\rightarrow 0.0578627940639531,\\[6pt]
&f[1,4,5]\rightarrow f[2,3,5], f[3,2,5]\rightarrow f[4,1,5]
\,.
\end{aligned}
\end{equation*}
\begin{equation*}
\Rc_4 = 0.05683770674055544 \,, \;  \Rc_6 = -0.8942359209665893 \,,
\end{equation*}
\begin{equation*}
\text{masses}^2 = (1.6346,0.33106,0.14162,-0.04085) \,, \;  \vec{v} = (0.48741,0.84003,0.20034,0.12901) \,.
\end{equation*}

\section{Numerical procedure}\label{ap:num}

To find de Sitter solutions of type IIB supergravity with intersecting $O_5/D_5$, a problem presented in section \ref{sec:framework}, we developed a numerical procedure in \emph{Wolfram Mathematica 12}. We commented on this procedure in section \ref{sec:dSsol}, and we now detail it in this appendix. The problem amounts to solving 46 scalar equations involving 43 variables, subject to the constraints \eqref{constraints}. We first rewrite each of the 46 equations in the form $E_i = 0 \,, \; i = 1, ...,  46$, and consider
\begin{equation}
s \equiv \sum_{i=1}^{46} E_i^2 \,.
\end{equation}
Then, $s$ vanishes at one of its minima if and only if all the equations are satisfied. One can thus look for solutions by using minimization with constraints algorithms, such as \code{NMinimize}. Without the constraints \eqref{constraints}, it is known that $s$ possesses multiple zeros that correspond for instance to Minkowski solutions, some of which we recovered as sanity checks (see section \ref{sec:Mink}). For de Sitter solutions, the roots of $s$ should then fall into the subspace of parameters obeying the constraints \eqref{constraints}.

The solutions considered here are numerical. Thus, for a given output of the minimization algorithm (denoted $\pb \in \mathbb{R}^{43}$), one needs criteria to decide whether it is likely to be an approximation of an exact solution. The most obvious criteria are the following:
\begin{itemize}
	\item The value of $s$ at the output $\pb$, denoted $\Delta \equiv s(\pb)$, should be very close to 0.
	\item One should check that each equation is solved separately to a good accuracy and compute the maximal error, $\varepsilon \equiv \max_{i} \lvert E_i(\pb) \vert$. This is the most relevant quantity, but note that it is strongly related to the previous one by $\Delta \sim \varepsilon^2$.
	\item The value of $\Rc_4$ at the output should not be too small, to ensure that we are not dealing with some numerical error around a Minkowski solution.
\end{itemize}
In practice, we restrict to outputs with $\Delta \sim \varepsilon^2 \sim 10^{-30}$. Indeed, we noticed a strong distinction between outputs with this level of accuracy and others, which were by far less precise. This level of precision is actually what we obtain while recovering known solutions of some simple equations: for instance one can minimize $s=(x^2-2)^2$ with the technique described below, and one gets $\bar{p} \approx \pm \sqrt{2}$ with $\Delta \sim 10^{-30}$. Here, the best accuracy obtained for a solution is $\Delta \sim 10^{-33}$. The value obtained for $\Rc_4$ would always fall into the interval $[10^{-3}, 10^{-1}]$ which is way larger than $\varepsilon$. Also, the values obtained for the variables are always in the range $[10^{-4}, 10]$, which is reasonably large compared to the magnitude of the numerical error. Finally, the accuracy obtained in a known no-go situation, e.g.~by setting $F_1=0$, is at best $\Delta \sim 10^{-10}$, allowing a clear distinction.

To obtain our solutions, we followed the procedure below:
\begin{itemize}
	\item First, we numerically minimize the function $s$ together with the constraints \eqref{constraints}, with \code{NMinimize}. To avoid the algorithm converging too easily towards Minkowski solutions, we actually minimize $s/\Rc_4^2$ instead. Regarding the parameters, we used the method \code{RandomSearch} with \code{SearchPoints $\rightarrow$ 50} and  \code{Tolerance $\rightarrow$ 0}. It is useful to set one variable to a fixed value to avoid \code{NMinimize} starting with initial points where $\Rc_4^2$ is close to 0 (potentially returning a $1/0$ error). We often set $T_{10}^1=10$. Note that this value can later be rescaled, as we do for instance with $\lambda$ in section \ref{sec:regime}. The restricted number of search points allows to obtain a first result $\pb_1$ in a short amount of time, which will be refined in the following steps.
	
	\item One can then look for a much more accurate solution, $\pb_2$, in a small region around $\pb_1$. To do that, we construct a small ball centered in $\pb_1$ with a radius $r=r_0 \times \sqrt{\Delta_1}$, with initially the constant $r_0 \sim \mathcal{O}(1)$, and we then tune it a posteriori to obtain the best accuracy. We then run a \code{FindMinimum} of $s$ restricted to this ball, now with high \code{AccuracyGoal $\rightarrow$ 100} and \code{PrecisionGoal $\rightarrow$ 100}. This is quasi-instantaneous. One can then adjust the radius by tuning $r_0$: a ball with larger radius more likely contains a good solution, but a too large radius might refrain the \code{FindMinimum} from converging at all.
	
	\item If the radius can easily be adjusted such that $\Delta_2 \sim 10^{-30}$ with $\Rc_4 > 10^{-3}$, then we consider $\pb_2$ to be a satisfactory solution.
\end{itemize}
We obtained this way 17 solutions as further described in section \ref{sec:dSsol}.

Finally, we note that there seems to be no solution with $T_{10}^2 = 0 = T_{10}^3$, even when one increases the number of initial points in \code{NMinimize} to \code{500}, and when one does not impose the orientifold projection along $(34)$ (see section \ref{sec:Op}), thus allowing for more variables. We also looked for solutions with a fixed number of non-zero structure constants, namely $1$ or $2$. For those cases there was no output of the \code{NMinimize} with $\Delta < 10^{-3}$, which makes it unlikely that solutions should be found around these points. In other words, it seems that there are no solutions with only $1$ or $2$ non-vanishing structure constants. We refer to section \ref{sec:dSsol} for more comments.

\section{Change of basis and algebra identification}\label{ap:alg}

In this appendix, we discuss changes of basis and further tools allowing us to identify the algebras underlying the 6d group manifolds in solutions 14 to 17, as discussed in section \ref{sec:compactness}. We start with solutions 16 and 17. Solution 17 has the following non-zero structure constants
\beq
\mbox{Solution 17:}\quad \f{1}{45}, \f{2}{35}, \f{2}{45}, \f{2}{46}, \f{3}{16}, \f{3}{25}, \f{4}{15}, \f{6}{14} \ , \label{sol17fabc}
\eeq
while solution 16 has the same set together with $\f{3}{15}$. The following change of basis
\beq
{e^{a\neq 2}}' = e^a \ , \ {e^2}'= e^2 + \frac{\f{3}{15}}{\f{3}{25}} e^1 \label{chgebasis1}
\eeq
leaves all structure constants of solution 16 invariant except
\beq
{\f{3}{15}}'=0 \ , \ {\f{2}{45}}'= \f{2}{45} + \frac{\f{1}{45}\f{3}{15}}{\f{3}{25}} \neq 0 \ ,
\eeq
where we verified ${\f{2}{45}}' \neq 0$ given the values in the solution. The change of basis \eqref{chgebasis1} thus brings the set of non-zero structure constants to \eqref{sol17fabc}, i.e.~that of solution 17.  From this set, one identifies the following nilradical, with non-zero structure constants
\beq
\mathfrak{n}=\{1,2,3,4,6\} \ , \ \f{2}{46}, \f{3}{16}, \f{6}{14} \ ,
\eeq
where the numbers in $\mathfrak{n}$ are the directions of the contributing algebra vectors. One verifies that this is a five-dimensional, indecomposable, two-step nilpotent algebra, identified as $\mathfrak{g}_{5.3}$ in \cite{Bock}. Since algebras generated from \eqref{sol17fabc} are unimodular and indecomposable (especially given their $\mathfrak{n}$ is five-dimensional and indecomposable), the identification of the nilradical implies that algebras of solutions 16 and 17 are among the two of table 27 of \cite{Bock}. According to Theorem 8.3.4 and the following remark there, both algebras of that table admit a lattice, so it is the case of our algebras as well. For completeness, we determine an isomorphism for each algebra of solutions 16 and 17 to the algebra $\mathfrak{g}_{6.76}^{-1}$ in table 27 of \cite{Bock}: this identifies them completely.\footnote{A first step in doing so is the following relabeling on the set \eqref{sol17fabc}
\beq
1\rightarrow 5,\ 2\rightarrow 3,\ 3\rightarrow 1,\ 4\rightarrow 4,\ 5\rightarrow 6,\ 6\rightarrow 2 \ ,\qquad \f{1}{25}, \f{1}{36}, \f{2}{45}, \f{3}{16}, \f{3}{24}, \f{3}{46}, \f{4}{56}, \f{5}{46}  \ . \nn
\eeq
The resulting set of structure constant is close to that in the table.} We also verify the absence of an isomorphism to the second algebra of that table.\\

We proceed similarly for solutions 14 and 15. Here are their structure constants and nilradical
\bea
& \mbox{Solution 14:}\quad \f{1}{45}, \f{1}{46}, \f{2}{35}, \f{2}{45}, \f{2}{46}, \f{3}{15}, \f{3}{25}, \f{6}{14} \ ,\quad \mathfrak{n}=\{1,2,3,6\} \ , \label{sol14fabc}\\
& \mbox{Solution 15:}\quad  \f{2}{35}, \f{2}{45}, \f{3}{15}, \f{3}{16}, \f{3}{25}, \f{4}{16}, \f{6}{14} \ ,\quad \mathfrak{n}=\{2,3,4,6\} \ ,  \label{sol15fabc}
\eea
and their nilradical is in both cases the four-dimensional abelian algebra, denoted $\mathfrak{n} = 4 \mathfrak{g}_1$ in \cite{Bock}. They are unimodular, and they look indecomposable, but given that their nilradical is decomposable, one may have a doubt. This point needs to be settled for the identification of the algebra. Let us first focus on solution 15: we can perform the following change of basis
\beq
{e^{a\neq {2,3}}}' = e^a \ , \ {e^2}'= e^2 + \frac{\f{3}{15}}{\f{3}{25}} e^1 \ ,\ {e^3}'= e^3 - \frac{\f{3}{16}}{\f{4}{16}} e^4 \label{chgebasis2}
\eeq
that leaves all structure constants invariant except
\beq
{\f{3}{15}}'={\f{3}{16}}'=0 \ , \ {\f{2}{45}}'= \f{2}{45} + \f{2}{35}\frac{\f{3}{16}}{\f{4}{16}} = 0 \ .
\eeq
The fact ${\f{2}{45}}' = 0$ comes at first sight as a surprise, and is obtained with the values of solution 15. It can be understood as due to the Jacobi identity along $2$, expressed as
\beq
\forall b,c,d,\ f^2{}_{e[b} f^e{}_{cd]}=0 \Leftrightarrow \f{2}{45}\f{4}{16} + \f{2}{35} \f{3}{16} = 0 \ \ \mbox{for the set}\ \eqref{sol15fabc} \ .
\eeq
This way, we are left with only four structure constants
\beq
\mbox{Solution 15}':\quad  {\f{2}{35}}', {\f{3}{25}}', {\f{4}{16}}', {\f{6}{14}}' \ , \label{sol15fabcprime}
\eeq
which form two pairs along $2,3,5$ and $1,4,6$. The algebra is then decomposable, into two three-dimensional solvable algebras, each of nilradical $2\mathfrak{g}_1$. Given the signs of the structure constants, they are both identified as $\mathfrak{g}_{3.4}^{-1}$ in \cite{Bock}. The latter admits a lattice, and so does the algebra $\mathfrak{g}_{3.4}^{-1} \oplus \mathfrak{g}_{3.4}^{-1}$ of our solution 15.

The same happens with our solution 14: we perform the change of basis
\beq
{e^{a\neq {2,3,6}}}' = e^a \ , \ {e^2}'= e^2 + \frac{\f{3}{15}}{\f{3}{25}} e^1 \ ,\ {e^3}'= e^3 + \frac{\f{1}{46}\f{2}{45}- \f{1}{45}\f{2}{46}}{\f{1}{46}\f{2}{35}} e^4 \ ,\ {e^6}'= e^6 + \frac{\f{1}{45}}{\f{1}{46}} e^5 \label{chgebasis3}
\eeq
where one verifies that the coefficient in front of $e^4$ is non-zero. This leaves all structure constants invariant except
\beq
{\f{1}{45}}'={\f{2}{45}}'={\f{3}{15}}'=0 \ , \ {\f{2}{46}}'= \f{2}{46} + \f{1}{46}\frac{\f{3}{15}}{\f{3}{25}} = 0 \ .
\eeq
Again the last annihilation can be verified on the values of solution and holds thanks to the Jacobi identity
\beq
\forall b,c,d,\ f^3{}_{e[b} f^e{}_{cd]}=0 \Leftrightarrow \f{3}{15}\f{1}{46} + \f{3}{25} \f{2}{46} = 0 \ \ \mbox{for the set}\ \eqref{sol14fabc} \ .
\eeq
We are left with only four structure constants
\beq
\mbox{Solution 14}':\quad  {\f{2}{35}}', {\f{3}{25}}', {\f{1}{46}}', {\f{6}{14}}' \ , \label{sol14fabcprime}
\eeq
which form two pairs along $2,3,5$ and $1,4,6$. As for solution 15, the algebra is then decomposable, into two three-dimensional solvable algebras, each of nilradical $2\mathfrak{g}_1$. The signs of the structure constants are different than those of solution 15, both sets are now identified as $\mathfrak{g}_{3.5}^{0}$ in \cite{Bock}. That algebra admits a lattice, and so does the algebra $\mathfrak{g}_{3.5}^{0} \oplus \mathfrak{g}_{3.5}^{0}$ of our solution 14. We conclude that for these 4 solutions, lattices could be found so their manifold $\mmm$ can be made compact, as further discussed in section \ref{sec:compactness}.

For completeness, we give here the new metric obtained for solutions 14 and 15 after the change of basis. With $e'=Me$ as in \eqref{chgebasis2} and \eqref{chgebasis3}, the metric goes from $\delta_{ab}$ to $g_{ab}$ with $g=M^{-\top}\, \delta\, M^{-1}$. For convenience, we write down $g^{-1}$: in both solutions, it is expressed in terms of the initial structure constants as follows
\bea
g^{-1}= &\, \left( \begin{array}{cccccc} 1 & g^{12} & & & & \\ g^{12} & 1+ (g^{12})^2 & & & & \\  & & 1 + (g^{34})^2 & g^{34} & & \\  & & g^{34} & 1 & & \\  & & & & 1 & g^{56} \\  & & & & g^{56} & 1+ (g^{56})^2  \end{array} \right) \ ,\label{g-1}\\
\mbox{Solution 14}'\!: &\quad g^{12}= \frac{\f{3}{15}}{\f{3}{25}} \ , \ g^{34}=\frac{\f{1}{46}\f{2}{45}- \f{1}{45}\f{2}{46}}{\f{1}{46}\f{2}{35}} \ , \ g^{56}= \frac{\f{1}{45}}{\f{1}{46}} \ ,\nn\\
\mbox{Solution 15}'\!: &\quad g^{12}= \frac{\f{3}{15}}{\f{3}{25}} \ , \ g^{34}= - \frac{\f{3}{16}}{\f{4}{16}} \ , \ g^{56}=0\ .  \nn
\eea
Note that the two changes of basis preserve the sources pairs of directions, i.e.~the pairs (12), (34), (56) remain among themselves. One even has $e^1 \w e^2 = {e^1}' \w {e^2}'$, etc., so the sources volume forms and contributions $T_{10}^I$ are unchanged. Similarly, the new metric is block diagonal in (12), (34), (56), each block remaining of determinant 1. This will be useful in \cite{Andriot:2020vlg}. We give explicitly in appendix \ref{ap:sol} the numerical solutions $14{}'$ and $15{}'$ obtained after the change of basis \eqref{chgebasis3} and \eqref{chgebasis2}; the flux components, as well as the other variables, are then expressed in the ${e^a}'$ basis.

\section{4d kinetic terms}\label{ap:4d}

In this appendix we compute on general grounds kinetic terms for 4d scalar fields, as in the 4d action \eqref{S4dgen}, obtained from certain fluctuations in a 10d theory. We consider the following 10d metric, with 4 and 6-dimensional diagonal blocks
\beq
\d s^2_{10} = \tau^{-2}(x)\, g_{\mu \nu}(x)\, \d x^{\mu} \d x^{\nu} + \rho(x)\, g_{mn}(x,y)\, \d y^m \d y^n\ ,\ \ \mu =0, \dots, 3,\; m=4, \dots, 9\ .\label{metric10}
\eeq
The metric $g_{\mu \nu}$ will eventually correspond to the 4d Einstein frame metric. For now, we compute the 10d Ricci scalar, with Levi-Civita connection
\bea
\R_{10} & = \tau^2 \R_4 + \rho^{-1} \R_6 - \nabla_{\mu} \left( 3 \tau^4 \del^{\mu} \tau^{-2} + \tau^2 \rho^{-1} g^{mn} \del^{\mu} (\rho g_{mn}) \right) \\
& - \frac{9}{2} \tau^6 (\del \tau^{-2})^2 - 2 \tau^4 \del_{\mu} \tau^{-2} \del^{\mu} (\rho g_{mn}) \rho^{-1} g^{mn} \nn\\
& - \frac{\tau^2}{4} \del_{\mu} (\rho g_{mn}) \rho^{-1} g^{mn} \del^{\mu} (\rho g_{pq}) \rho^{-1} g^{pq} + \frac{\tau^2}{4} \del_{\mu} (\rho g_{mn}) \del^{\mu} (\rho^{-1} g^{mn}) \nn\ ,
\eea
where $g_{\mu\nu}$ is used to define covariant derivatives, lift indices, and in the squares. Here and in the rest of this appendix, $\R_6 $ denotes the Ricci scalar for $g_{mn}$ in \eqref{metric10}, with purely 6d derivatives. We already factored out the $\rho$-dependency, so it will eventually correspond to $\R_6(\sigma_1,\sigma_2)$ in \eqref{R6sigma}. We now perform an integration by parts of the previous total derivative, with a prefactor that will be justified below. In particular, the derivative being 4d, purely 6d dependent quantities denoted in the integral by the dots do not matter. With $g_4={\rm det}\, g_{\mu\nu}$ we get
\bea
& \int \d^4 x \sqrt{|g_4|} \tau^{-2} \dots \R_{10} \label{int1}\\
= & \int \d^4 x \sqrt{|g_4|} \dots \Big( \R_4 + \tau^{-2} \rho^{-1} \R_6 - \frac{3}{2} \tau^4 (\del \tau^{-2})^2 - \tau^2 \del_{\mu} \tau^{-2} \del^{\mu} (\rho g_{mn}) \rho^{-1} g^{mn} \nn\\
& \phantom{\int \d^4 x \sqrt{|g_4|} \dots ( } - \frac{1}{4} \del_{\mu} (\rho g_{mn}) \rho^{-1} g^{mn} \del^{\mu} (\rho g_{pq}) \rho^{-1} g^{pq} + \frac{1}{4} \del_{\mu} (\rho g_{mn}) \del^{\mu} (\rho^{-1} g^{mn}) \Big) \nn\ .
\eea
For any invertible matrix $A$
\beq
\del_{\mu} \ln {\rm det} A = {\rm Tr} (A^{-1} \del_{\mu} A) \ .
\eeq
Therefore, if ${\rm det}\, g_{mn} = g_6$ is independent of $x^{\nu}$, one deduces $g^{mn} \del_{\mu} g_{mn} = 0$. This simplifies drastically the previous expression towards
\bea
& \int \d^4 x \sqrt{|g_4|} \tau^{-2} \dots \R_{10} \label{intindpt}\\
= & \int \d^4 x \sqrt{|g_4|} \dots \Big( \R_4 + \tau^{-2} \rho^{-1} \R_6 - \frac{3}{2} \tau^4 (\del \tau^{-2})^2 - 6 \tau^2 \rho^{-1} \del_{\mu} \tau^{-2} \del^{\mu} \rho \nn\\
& \phantom{\int \d^4 x \sqrt{|g_4|} \dots (} - 9 \rho^{-2} (\del \rho )^2 - \frac{3}{2} \rho^{-2} (\del \rho)^2 + \frac{1}{4} \del_{\mu} (g_{mn}) \del^{\mu} (g^{mn}) \Big) \nn\ .
\eea

We now provide some context. We start with the 10d action
\beq
{\cal S} = \frac{1}{2 \kappa_{10}^2} \int \d^{10} x \sqrt{|g_{10}|} e^{-2\phi} \Big( \R_{10} + 4 (\del \phi)_{10}^2 \Big) \ ,
\eeq
where the square is made with the 10d metric, and $\kappa_{10}$ is a constant. We now take $e^{\phi} = e^{\phi_0(y)} e^{\delta \phi (x)}$ and $\tau = e^{-\delta \phi} \rho^{\frac{3}{2}}$. With the metric \eqref{metric10}, the previous action first becomes
\beq
{\cal S} = \frac{1}{2 \kappa_{10}^2} \int \d^4 x \sqrt{|g_{4}|} \tau^{-2} \int \d^6 y \sqrt{|g_{6}|} e^{-2\phi_0} \Big( \R_{10} + 4 (\del \phi)_{10}^2 \Big) \ .
\eeq
If we now restrict to the case where $g_6$ is independent of $x^{\mu}$, this justifies the integral prefactor of \eqref{int1}, and we can use the result of the computation \eqref{intindpt}. In addition, one computes
\beq
4 (\del \phi)_{10}^2 = 4 (\del \phi_0)_{6}^2 + 9 \rho^{-2} (\del \rho)^2 + \tau^4 (\del \tau^{-2})^2 + 6 \tau^2 \rho^{-1} \del_{\mu} \tau^{-2} \del^{\mu} \rho \ .
\eeq
Combined with \eqref{intindpt}, we finally obtain
\bea
{\cal S} = \frac{1}{2 \kappa_{10}^2} \int \d^4 x \sqrt{|g_{4}|} \int \d^6 y \sqrt{|g_{6}|} e^{-2\phi_0} \Big( & \R_4 + \tau^{-2} \rho^{-1} \R_6 + 4 (\del \phi_0)_{6}^2 \\
& - \frac{1}{2} \tau^4 (\del \tau^{-2})^2
- \frac{3}{2} \rho^{-2} (\del \rho)^2 + \frac{1}{4} \del_{\mu} (g_{mn}) \del^{\mu} (g^{mn}) \Big) \ . \nn
\eea
Introducing
\bea
& \hat{\tau}= \sqrt{2}\, M_p\, \ln \tau \, ,\ \hat{\rho}= \sqrt{\frac{3}{2}}\, M_p\, \ln \rho \, ,\quad M_p^2= \frac{1}{\kappa_{10}^2} \int \d^6 y \sqrt{|g_{6}|} e^{-2\phi_0}\,,\\
& V=  \frac{1}{2\kappa_{10}^2} \int \d^6 y \sqrt{|g_{6}|} e^{-2\phi_0} \Big(-\tau^{-2} \rho^{-1} \R_6 - 4 (\del \phi_0)_{6}^2  \Big) \ ,
\eea
we rewrite the above in the form of the 4d action \eqref{S4dgen}, namely
\beq
{\cal S} = \int \d^4 x \sqrt{|g_{4}|} \left( \frac{M_p^2}{2} \R_4  - \frac{1}{2} \left( (\del \hat{\tau})^2
+ (\del \hat{\rho})^2 - \frac{1}{4}\, M_p^2\, \del_{\mu} (g_{mn}) \del^{\mu} (g^{mn}) \right) - V \right) \ .
\eeq
This formula, especially the $\frac{1}{4} \del_{\mu} (g_{mn}) \del^{\mu} (g^{mn})$, matches the one used in appendix B of \cite{Andriot:2019wrs} to compute the kinetic term of the third scalar field $\sigma$; we will use it here again for several $\sigma^I$. Also, we match the conventions used in \cite{Andriot:2018ept, Andriot:2019wrs}, especially to derive the potential $V$, up to a redefinition of the 4d reduced Planck mass $M_p^2 = 2 M_{4 \, {\rm (there)}}^2$. While $V$ is the same, the reduced potential $\tilde{V}=V/M_{4}^2$ used there is then altered by a factor of 2.

We now compute the kinetic terms for two fields $\sigma_{I=1,2}(x)$, as e.g.~those of section \ref{sec:pot}. They are defined as follows
\bea
g_{mn} \d y^m \d y^n  = &\ \sigma_1^A \sigma_2^B \delta_{ab} e^a e^b + \sigma_1^A \sigma_2^A \delta_{cd} e^c e^d +\sigma_1^B \sigma_2^A \delta_{ef} e^e e^f +\sigma_1^B \sigma_2^B \delta_{gh} e^g e^h \ ,\\
{\rm where}\ & a,b=1,\dots, p-3 - N_0\, ,\ c,d=p-3 - N_0+1,\dots, p-3\, ,\nn\\
& e,f=p-3+1, \dots, 2(p-3)-N_0\, ,\ g,h=2(p-3)-N_0+1, \dots, 6 \ ,\nn
\eea
i.e.~$\sigma_I^A$ is along $p-3$ directions and $\sigma_I^B$ along $9-p$ ones, and the related sources $1$ and $2$ share $N_0$ common parallel directions. One has $e^a= e^a{}_m \d y^m$ where the vielbein $e^a{}_m(y)$ does not depend on 4d coordinates. In addition, the powers, $A=p-9$, $B=p-3$, verify $A(p-3)+B(9-p)=0$. This allows to have $g_6$ independent of $x^{\mu}$. This is most easily seen by introducing the diagonal matrix $M$ (we indicate in subscript the size of the blocks)
\bea
& g_{mn} \d y^m \d y^n = M_{ab} e^a e^b \ ,\\
& M={\rm diag}\Big(\{\sigma_1^A \sigma_2^B \}_{p-3-N_0}, \{\sigma_1^A \sigma_2^A \}_{N_0},\ \{\sigma_1^B \sigma_2^A \}_{p-3-N_0},\ \{\sigma_1^B \sigma_2^B \}_{9-p-(p-3- N_0)} \Big) \ ,\nn\\
& g_6 = ({\det}\, e)^2 {\det}\, M = ({\det}\, e)^2 (\sigma_1 \sigma_2)^{A(p-3)+B(9-p)} = ({\det}\, e)^2 \ .\nn
\eea
The kinetic terms for the $\sigma_I$ is then given as above by
\bea
- \frac{1}{4} \del_{\mu} (g_{mn}) \del^{\mu} (g^{mn}) & = - \frac{1}{4} \del_{\mu} (M_{ab}) \del^{\mu} (M^{ab}) \\
& = \frac{1}{4} \left( (-6AB) (\sigma_1^{-2} (\del \sigma_1 )^2 + \sigma_2^{-2} (\del \sigma_2 )^2) - 12 (B^2-6N_0) (\sigma_1 \sigma_2 )^{-1} \del \sigma_1 \del \sigma_2   \right) \nn\\
& = \frac{3}{4} \left( (-AB + B^2-6N_0) \big(\del \ln \frac{\sigma_1}{\sigma_2} \big)^2 + (-AB -( B^2-6N_0)) \big(\del \ln (\sigma_1 \sigma_2) \big)^2  \right) \ . \nn
\eea
It is easy from the last line to define canonical fields $\hat{\sigma}_I$. One also reproduces the result of \cite{Andriot:2019wrs} for one $\sigma$ by setting $\sigma_2=1$. Applying this result to the case of this paper, i.e.~$p=5$ and $N_0=0$, we obtain
\bea
{\cal S} = \int \d^4 x \sqrt{|g_{4}|} &\, \Bigg( \frac{M_p^2}{2} \R_4 - V \\
&\,  - \frac{1}{2} \Big( (\del \hat{\tau})^2 + (\del \hat{\rho})^2 + 12\, M_p^2\, \left( (\del \ln \sigma_1 )^2 + (\del \ln \sigma_2 )^2 - \del \ln \sigma_1 \del \ln \sigma_2 \right) \Big)  \Bigg) \ .\nn
\eea
We rewrite this formula in \eqref{S4dkin1}.\\

Finally, the fluctuation of the 6d Ricci scalar contributes through six terms to the scalar potential $V$ as in \eqref{R6sigma}. We give here the explicit contributions:
\begin{align}
-2 R_1 &= \left(\f{1}{35} \right)^2 + \left(\f{1}{36} \right)^2 + \left(\f{1}{45} \right)^2 + \left(\f{1}{46} \right)^2 + \left(\f{2}{35} \right)^2 + \left(\f{2}{36} \right)^2 + \left(\f{2}{45} \right)^2 + \left(\f{2}{46} \right)^2 \,, \nn \\[6pt]
-2 R_2 &= \left(\f{3}{15} \right)^2 + \left(\f{3}{16} \right)^2 + \left(\f{3}{25} \right)^2 + \left(\f{3}{26} \right)^2 + \left(\f{4}{15} \right)^2 + \left(\f{4}{16} \right)^2 + \left(\f{4}{25} \right)^2 + \left(\f{4}{26} \right)^2 \,, \nn \\[6pt]
-2 R_3 &= \left(\f{5}{13} \right)^2 + \left(\f{35}{14} \right)^2 + \left(\f{5}{23} \right)^2 + \left(\f{5}{24} \right)^2 + \left(\f{6}{13} \right)^2 + \left(\f{6}{14} \right)^2 + \left(\f{6}{23} \right)^2 + \left(\f{6}{24} \right)^2 \,, \nn \\[6pt]
-R_4 &= \f{1}{35} \, \f{3}{15} + \f{1}{36} \, \f{3}{16} + \f{2}{35} \, \f{3}{25} + \f{2}{36}  \, \f{3}{26} \nn \\
&+\f{1}{45} \, \f{4}{15} + \f{1}{46} \, \f{4}{16} + \f{2}{45} \, \f{4}{25} + \f{2}{46} \, \f{4}{26} \,, \label{coeff} \\[6pt]
-R_5 &= \f{3}{51} \, \f{5}{31} + \f{4}{51} \, \f{5}{41} + \f{3}{52} \, \f{5}{32} + \f{4}{52}  \, \f{5}{42} \nn \\
&+\f{3}{61} \, \f{6}{31} + \f{4}{61} \, \f{6}{41} + \f{3}{62} \, \f{6}{32} + \f{4}{62} \, \f{6}{42} \,, \nn \\[6pt]
-R_6 &= \f{5}{13} \, \f{1}{53} + \f{5}{14} \, \f{1}{54} + \f{5}{23} \, \f{2}{53} + \f{5}{24}  \, \f{2}{54} \nn \\
&+\f{6}{13} \, \f{1}{63} + \f{6}{14} \, \f{1}{64} + \f{6}{23} \, \f{2}{63} + \f{6}{24} \, \f{2}{64} \,, \nn
\end{align}
where we drop the background label ${}^0$ for simplicity.

\end{appendix}

\newpage

\providecommand{\href}[2]{#2}\begingroup\raggedright
\endgroup

\end{document}